\documentclass{pasj00}

\usepackage{graphics}
\usepackage{natbib}
\usepackage{url}
\usepackage{lscape}
\usepackage{deluxetable}
\usepackage{multirow}

\newcommand{\udots}{$\cdot\cdot\cdot$}

\begin{document}
\SetRunningHead{M. Hayashi, et al.}{Physical conditions of the ISM at $z\sim$1.5}

\title{Physical conditions of the interstellar medium in star-forming galaxies at $z\sim$1.5}

\author{%
   Masao \textsc{Hayashi},\altaffilmark{1}
   Chun \textsc{Ly},\altaffilmark{2,10}
   Kazuhiro \textsc{Shimasaku},\altaffilmark{3,4}
   Kentaro \textsc{Motohara},\altaffilmark{5}
   Matthew A. \textsc{Malkan},\altaffilmark{6}
   Tohru \textsc{Nagao},\altaffilmark{7}
   Nobunari \textsc{Kashikawa},\altaffilmark{1,8}
   Ryosuke \textsc{Goto},\altaffilmark{3}
   and
   Yoshiaki \textsc{Naito}\altaffilmark{9}
}
\altaffiltext{1}{Optical and Infrared Astronomy Division, National Astronomical Observatory, 2-21-1 Osawa, Mitaka, Tokyo 181-8588}
\email{masao.hayashi@nao.ac.jp}
\altaffiltext{2}{NASA Goddard Space Flight Center, 8800 Greenbelt Road, Greenbelt, MD 20771, U.S.A.}
\altaffiltext{3}{Department of Astronomy, Graduate School of Science, The University of Tokyo, Hongo, Tokyo 113-0033}
\altaffiltext{4}{Research Center for the Early Universe, The University of Tokyo, Hongo, Tokyo 113-0033}
\altaffiltext{5}{Institute of Astronomy, Graduate School of Science, The University of Tokyo, Mitaka, Tokyo 181-0015}
\altaffiltext{6}{Department of Physics and Astronomy, University of California at Los Angeles, Los Angeles, CA 90095-1547, U.S.A.}
\altaffiltext{7}{Research Center for Space and Cosmic Evolution, Ehime University, 2-5 Bunkyo-cho, Matsuyama, Ehime 790-8577}
\altaffiltext{8}{Department of Astronomy, School of Science, Graduate University for Advanced Studies, Mitaka, Tokyo 181-8588}
\altaffiltext{9}{Institute for Cosmic Ray Research, The University of Tokyo, 5-1-5 Kashiwanoha, Kashiwa, Chiba 277-8582}
\altaffiltext{10}{NASA Postdoctoral Fellow}


\maketitle

\begin{abstract}
We present results from Subaru/FMOS near-infrared (NIR) spectroscopy
of 118 star-forming galaxies at $z\sim1.5$ in the Subaru Deep Field.  
These galaxies are selected as [O\emissiontype{II}]$\lambda$3727
emitters at $z\approx$ 1.47 and 1.62 from narrow-band imaging. We
detect H$\alpha$ emission line in 115 galaxies,
[O\emissiontype{III}]$\lambda$5007 emission line in 45 galaxies, and
H$\beta$, [N\emissiontype{II}]$\lambda$6584, and
[S\emissiontype{II}]$\lambda\lambda$6716,6731 in 13, 16,
and 6 galaxies, respectively. Including the [O\emissiontype{II}]
emission line, we use the six strong nebular emission lines in the
individual and composite rest-frame optical spectra to investigate
physical conditions of the interstellar medium in star-forming
galaxies at $z\sim$1.5.     
We find a tight correlation between H$\alpha$ and
[O\emissiontype{II}], which suggests that [O\emissiontype{II}] can be
a good star formation rate (SFR) indicator for galaxies at $z\sim1.5$.  
The line ratios of H$\alpha$/[O\emissiontype{II}] are consistent with
those of local galaxies. 
We also find that [O\emissiontype{II}] emitters have strong
[O\emissiontype{III}] emission lines. The
[O\emissiontype{III}]/[O\emissiontype{II}] ratios are larger than
normal star-forming galaxies in the local Universe, suggesting a
higher ionization parameter. Less massive galaxies have larger
[O\emissiontype{III}]/[O\emissiontype{II}] ratios. With evidence that
the electron density is consistent with local galaxies, the high
ionization of galaxies at high redshifts may be attributed
to a harder radiation field by a young stellar population and/or an
increase in the number of ionizing photons from each massive star. 
\end{abstract}

\section{Introduction}

Recent studies indicate that
H\emissiontype{II} regions in star-forming galaxies at $z\approx$ 1--3
differ from those in local galaxies
\citep[e.g.,][]{Brinchmann2008,Liu2008,Shirazi2014,Newman2014}. 
The measured ionization parameter in star-forming regions at high
redshifts is found to be higher than in normal star-forming galaxies in
the local Universe  
\citep{Nakajima2013,Richardson2013,Amorin2014,Holden2014,Ly2014,Nakajima2014,Shirazi2014}.
It is also suggested that H\emissiontype{II} regions of high-$z$
galaxies have a harder stellar ionization radiation field
\citep{Steidel2014} and higher electron densities
\citep{Hainline2009,Bian2010,Shirazi2014,Shimakawa2014c}.   

The SFR of galaxies is known to increase by $\sim30$ times from the
local Universe up to $z\approx2.5$ at a given stellar mass  
\citep[e.g.,][]{Daddi2007,Ly2011,Whitaker2014}, and the number density of active
galactic nuclei (AGNs) peaks at $z\approx$ 2--3 \citep{Hopkins2007}. Moreover, studies
suggest that the gas fraction in galaxies increases up to $\sim50$\%
at $z\sim2$ from $\lesssim$10\% at $z\sim0$ \citep{Leroy2008,Tacconi2010,Geach2011}. 
Given these facts, it is not surprising that AGN and supernova
feedback, and a larger contribution of young massive stars to the radiation
field can alter the conditions of the interstellar matter (ISM) in high-$z$
galaxies.   

Many studies of the metal content in
H\emissiontype{II} regions of galaxies, and its redshift evolution have
revealed a stellar mass-metallicity relation:  more
massive star-forming galaxies are more chemically enriched.
This correlation evolves
toward lower metallicity (at a given stellar mass) at higher redshift 
\citep[e.g.,][]{Tremonti2004,Savaglio2005,Shapley2005,Erb2006a,Maiolino2008,Liu2008,hayashi2009,Queyrel2009,Queyrel2012,Yuan2009,Richard2011,Rigby2011,Yabe2012,Yabe2014,Roseboom2012,Stott2013,Yuan2013,Henry2013L,Zahid2013,Zahid2014,Wuyts2014,Divoy2014,Ly2014,Ly2014b,delosReyes2015}.    
However, the redshift dependence of the ISM in star-forming
galaxies can produce one of the major uncertainties in
metallicity measurements at high redshifts.       
This is because strong nebular emission line diagnostics
generally used to estimate the oxygen abundance of high-$z$ galaxies
are calibrated with metallicity derived by photoionization models or
metallicity based on electron temperature ($T_e$) in the local Universe
\citep[e.g.,][]{Pagel1979,McGaugh1991,Kewley2002,Pettini2004,Nagao2006,Kewley2008}.  
Metal emission lines, such as [O\emissiontype{II}] and
[O\emissiontype{III}] in H\emissiontype{II} regions, arise from
collisional excitation of ions with electrons. 
The intensity ratios of nebular emission lines are sensitive to not only
metallicity, but also the ionization state and electron density of the ISM.  
Using diagnostics that are calibrated with local galaxies requires that
we assume that the physical state of the gas in high-$z$ star-forming
galaxies is similar to that of local galaxies. 
However, recent studies are raising doubts that this
assumption holds at high redshifts.

In this paper, we focus on $z\sim1.5$ star-forming galaxies
and investigate the physical state of their ISM and how it compares to
local galaxies using six strong nebular emission lines
seen in the rest-frame optical wavelength:
[O\emissiontype{II}]$\lambda$3727,
H$\beta$,
[O\emissiontype{III}]$\lambda\lambda$4959,5007,
H$\alpha$,
[N\emissiontype{II}]$\lambda$6584,
and [S\emissiontype{II}]$\lambda\lambda$6716,6731.
[O\emissiontype{II}] and [S\emissiontype{II}] lines are essential to
derive the ionization parameter and electron density of ISM.
We obtained NIR spectroscopy for 118 star-forming
galaxies emitting an [O\emissiontype{II}]$\lambda$3727 line (hereafter
[O\emissiontype{II}] emitters) at $z\sim1.5$. Our targets are
extracted from two samples of [O\emissiontype{II}] emitters at
$z\approx$ 1.47 and 1.62 in the Subaru Deep Field
\citep[SDF;][]{Kashikawa2004}, which are identified by NB921
($\lambda_c=$9196\AA, $\Delta\lambda$=132\AA) or NB973
($\lambda_c=$9755\AA, $\Delta\lambda$=202\AA) narrow-band imaging
\citep{Ly2007,Ly2012} with Suprime-Cam \citep{Miyazaki2002} on the
Subaru Telescope.       

While previous studies at $z\approx1.5$--2.5 have begun to increase
spectroscopic samples, most of them only have observations based on a
limited number of nebular emission lines.
For example, four lines of H$\alpha$, H$\beta$, [O\emissiontype{III}]
and [N\emissiontype{II}] are measured for a few hundred galaxies at
$z\sim$1.4--1.6 in several deep fields
\citep{Stott2013,Zahid2014,Yabe2014}. \citet{Steidel2014} study 
179 star-forming galaxies at $z\sim2.3$. However, in these studies,
limited [O\emissiontype{II}] information is available.  
In contrast, all of our galaxies have the [O\emissiontype{II}]
luminosity measured by the narrow-band imaging.   
There are a few studies with the full rest-frame
optical spectra covering [O\emissiontype{II}], H$\beta$, [O\emissiontype{III}],
H$\alpha$, and [N\emissiontype{II}]. Most of the samples are not
adequate to discuss the difference and diversity statistically and/or
are biased toward massive galaxies or lensed galaxies (at most
$\sim$30 galaxies at $z\gtrsim1.5$ in \citet{Masters2014} and
\citet{Maier2014}). \citet{Shapley2014} have used a large sample of
118 galaxies, but it is $H$-band selected (i.e., mass-selected)
sample, and probes $z\sim2.3$, compared with our samples being
[O\emissiontype{II}]-selected at $z\sim1.5$ \citep{Ly2012}.   

The outline of this paper is as follows. The NIR spectroscopic
observations and the data reduction are described in \S
\ref{sec:OBSandDATA}. Identification of emission lines in the 
individual spectra and the measurement of emission-line fluxes are also
described. In \S \ref{sec:physical_state}, we study the physical state
of ISM in galaxies at $z\sim1.5$, focusing on measurements of the
electron density, dust extinction, AGN contribution, ionization parameter
and metal abundance.    
We then investigate the relationship between H$\alpha$ and
[O\emissiontype{II}] fluxes for individual galaxies, both of which are
widely used as star formation indicators \citep[e.g.,][]{Kennicutt1998}. 
In \S \ref{sec:discussion}, we discuss possible mechanisms causing any
differences between [O\emissiontype{II}] emitters at $z\sim1.5$ and
star-forming galaxies in the local Universe. 
We also compare our results with the previous studies on star-forming
galaxies at similar redshifts.  
In \S \ref{sec:conclusions}, concluding remarks are provided. All
magnitudes are provided on the AB system \citep{Oke1983}, and a
cosmology consisting of $h=0.7$, $\Omega_{m}=0.3$, and
$\Omega_{\Lambda}=0.7$ is adopted throughout this paper.

\section{Observations and Data}
\label{sec:OBSandDATA}

\begin{figure}
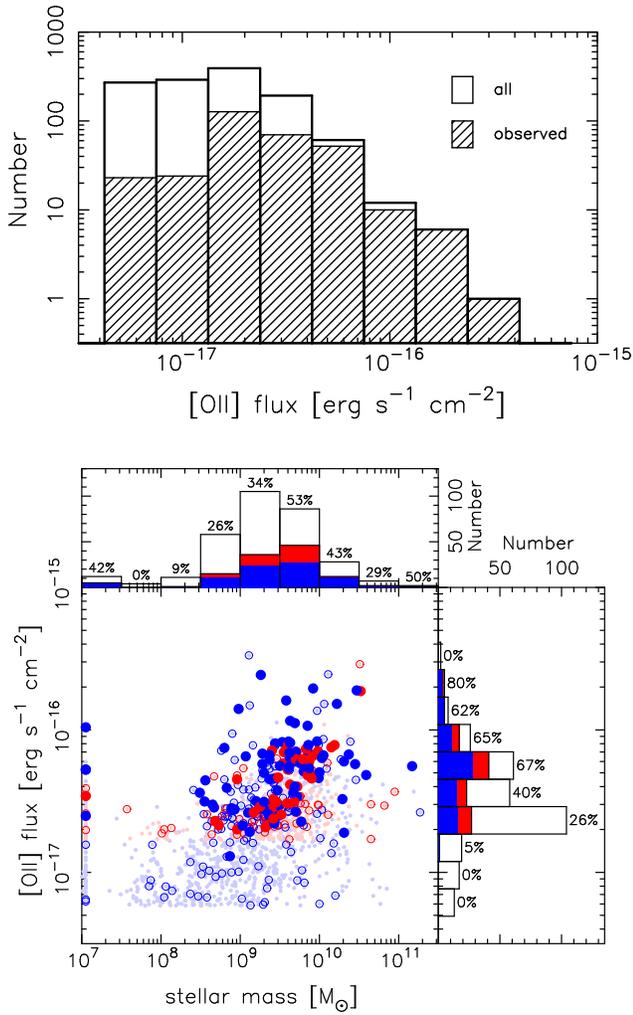

 \begin{center}
   \vspace{-2.0cm}
   \includegraphics[width=9.2cm]{./fig1a.eps} \\[-2mm]
   \hspace{-5mm}
   \includegraphics[width=7.9cm]{./fig1b.eps} 
   \vspace{-1.0cm}
 \end{center}
\caption{
{\it Top}: Histograms of [O\emissiontype{II}] flux measured from
narrow-band imaging for all of the NB921 and NB973
[O\emissiontype{II}] emitters at $z\sim1.5$ (open) and the emitters
targeted in our Subaru/FMOS observations (hatched).
{\it Bottom}: The distribution of [O\emissiontype{II}] flux measured from
narrow-band imaging as a function of stellar mass for
[O\emissiontype{II}] emitters at $z\sim1.5$.
The NB921 [O\emissiontype{II}] emitters at $z\approx1.47$ are shown in
blue, while the NB973 [O\emissiontype{II}] emitters at $z\approx1.62$
are shown in red. The circles in pale colors show all of the
[O\emissiontype{II}] emitters at $z\sim1.5$. The large circles show
the targeted 313 [O\emissiontype{II}] emitters in this study. Among
them, the filled ones show the 118 [O\emissiontype{II}] 
emitters with at least H$\alpha$ or [O\emissiontype{III}]
detected. The histograms on the right show the distribution of
[O\emissiontype{II}] flux for the targets (open) and the
spectroscopically confirmed sample (filled). The numbers signify the
percentage of the confirmed galaxies out of the observed ones in each
bin of [O\emissiontype{II}] flux. The histograms on the top show the
distribution of stellar mass of the targets and confirmed
galaxies. The galaxies for which the stellar masses are poorly
estimated are plotted as galaxies with stellar mass of $10^{7}$ \MO. 
}
\label{fig:target}
\end{figure}

\subsection{Star-forming galaxies at $z\sim1.5$}

\subsubsection{Observations}

Our samples in the SDF (\timeform{13h24m38.9s},
\timeform{+27d29m25.9s}) contain 933 NB921
[O\emissiontype{II}] emitters at $z\approx$ 1.450--1.485 with
[O\emissiontype{II}] flux $>$ 5.8 $\times10^{-18}$ erg s$^{-1}$
cm$^{-2}$, as well as, 328 NB973 [O\emissiontype{II}] emitters at
$z\approx$ 1.591--1.644 with [O\emissiontype{II}] flux $>$ 1.7 $\times10^{-17}$
erg s$^{-1}$ cm$^{-2}$ \citep{Ly2007,Ly2012}. 
Here, we briefly describe the selection of the [O\emissiontype{II}]
emitters and the measurement of [O\emissiontype{II}] fluxes.
We refer readers to \citet{Ly2007,Ly2012} for details.  
The emission-line galaxies are selected to have a color excess in
$z'-$NB (NB921 or NB973), where the $z'-$NB colors are corrected with
$i-z'$ color to properly estimate stellar continuum flux at the
wavelengths of the NB filters. Then, selections with $BR_ci'$
and $R_ci'z'$ color-color diagrams are adopted to identify the
[O\emissiontype{II}] emitters at $z\sim1.47$ or $1.62$.  
The [O\emissiontype{II}] fluxes ($F_{\rm [O\emissiontype{II}]}$) are
measured as follows:
\begin{equation}
F_{\rm [O\emissiontype{II}]} = \Delta{\rm NB} \frac{f_{\rm NB}-f_{z'}}{1-(\Delta{\rm NB}/\Delta z')},
\end{equation}
where $f_X$ and $\Delta{X}$ are the flux density and full width at half
maximum (FWHM) of bandpass ``$X$'', respectively. The FWHM of $z'$ is 955\AA.

To detect the six major nebular emission lines in rest-frame optical
for these [O\emissiontype{II}] emitters at $z\sim1.5$, NIR
spectroscopy was conducted with the Fiber Multi Object Spectrograph
\citep[FMOS;][]{Kimura2010} on the Subaru Telescope. FMOS is a unique
NIR spectrograph with 400 fibers over a wide field-of-view (FoV), 30
arcmin (diameter). 
The observations consist of two programs (S12A-026 and
S14A-018, PI: M. Hayashi). The first run was carried out on 7--9 April
2012 to observe 313 galaxies at $z\sim1.5$ (211 NB921 and 102 NB973
[O\emissiontype{II}] emitters) in high resolution mode in ``$H$-long''
(1.60--1.80$\mu$m), using three fiber configuration setups. 
We give a higher priority to galaxies with larger [O\emissiontype{II}]
flux (Figure \ref{fig:target}).
Six G- or K-type stars with $H=15.5$--17.3 mag are also included in
each setup as calibration stars.  
We use the higher spectral resolution mode ($R\sim2200$), because it
has higher system throughput and is more effective at avoiding
contamination from strong OH sky-lines than the low resolution
mode ($R\sim600$).
The $H$-band spectra enable us to observe H$\alpha$,
[N\emissiontype{II}]$\lambda$6584, and
[S\emissiontype{II}]$\lambda\lambda$6716,6731.     
The cross beam switching (CBS) method is adopted, where half of the
available fibers are allocated to the targets and, at the same time,
the other half of them observe nearby regions of blank sky. That is, a
pair of fibers (called ``A'' and ``B'') are allocated to a galaxy, and
ABAB dithering is performed for optimal sky subtraction.
Observations were only conducted during the last 1.5 nights, because
the weather was cloudy for the first one and half nights on 7--9
April. 
The total integration time was 315, 165, or 150 minutes for 143, 10, or
160 galaxies, respectively.   
The typical seeing was 0.8--1.0 arcsec in the $R$-band, which was measured
for the bright stars near the center of the FoV.    

\begin{figure*}
  \hspace{1.1cm}
  \begin{center}
    \includegraphics[width=17.5cm]{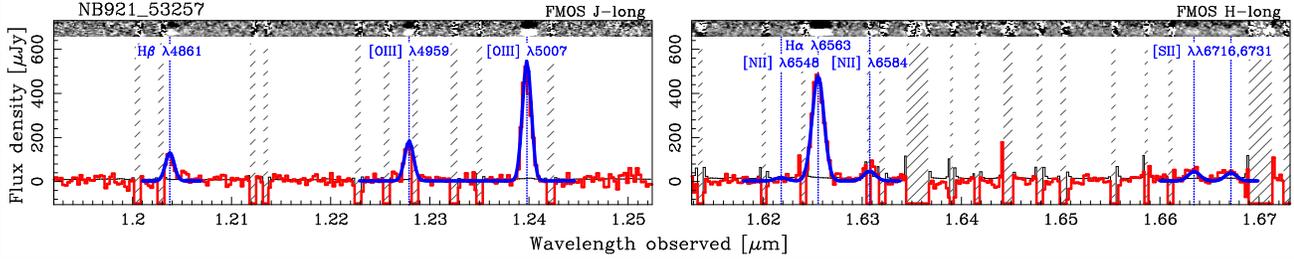} 
  \end{center}
\caption{
The spectra in $J$- (left) and $H$-band (right) taken with FMOS for
one of the brightest [O\emissiontype{II}] emitters.  
The red line shows the spectrum of the galaxy, and blue lines show the
Gaussian profile fits (\S \ref{sec:Gaussian_fit}).
The blue dotted lines show the wavelength of individual emission lines.
The black line shows the noise level. The hatched regions show the
areas that are masked for strong OH sky lines. 
The 2-D spectra in A and B fibers are also shown on top.    
Seven nebular emission lines, H$\beta$,
[O\emissiontype{III}]$\lambda\lambda$4959,5007, H$\alpha$,
[N\emissiontype{II}]$\lambda$6584, and
[S\emissiontype{II}]$\lambda\lambda$6716,6731, are seen.  
}
\label{fig:individual_spectrum}
\end{figure*}

The second run was carried out on 11--12 April 2014 to observe 125
galaxies at $z\sim1.5$ (81 NB921 and 44 NB973 [O\emissiontype{II}]
emitters) in high-resolution mode in ``$J$-long'' (1.11--1.35$\mu$m)
using two fiber configuration setups. The $J$-band spectra enable us
to observe H$\beta$ and
[O\emissiontype{III}]$\lambda\lambda$4959,5007.  
In the second run, we targeted galaxies observed in the first run
for spectral coverage of all emission lines. We
gave a higher priority to the galaxies with detections of H$\alpha$ in
the $H$-band spectrum. Moreover, since the $J$ and $H$ spectra were
taken in different observing conditions on different dates, we
allocate fibers to the same calibration stars for accurate flux
calibration in all observation runs. 
The CBS method is also adopted in the second run. On 11 April, almost
all of the night was affected by thin cirrus coverage, and the 
seeing was 0.7--1.1 arcsec. The sky cleared on 12 April, however,
the seeing was poor for the first half of the night ($\sim$1.4--1.7
arcsec). Thus, we only obtained optimal data under clear sky condition  
and sub-arcsec seeing (0.7--1.0 arcsec) for one half-night on 12
April. 
The total integration time was dependent on the target; 540 (390)
minutes for 49 (76) galaxies, respectively. 
Note that the data taken under marginal conditions are
also used in this study after correcting for the loss of flux.     

Figure \ref{fig:target} shows the observed [O\emissiontype{II}] flux
of the 313 [O\emissiontype{II}] emitters targeted in the FMOS
observations as a function of stellar mass (\S
\ref{sec:SED_fitting}). This figure shows that 
although our target sample is biased toward bright galaxies with
$F_{\rm [O\emissiontype{II}]}\gtrsim 5\times10^{-17}$ erg s$^{-1}$
cm$^{-2}$, compared with the whole sample of [O\emissiontype{II}]
emitters at $z\sim1.5$, 
we have observed star-forming galaxies with stellar mass covering
about three orders of magnitude.

\vspace{1em}
\subsubsection{Data reduction}

The FIBRE-pac data reduction package \citep{iwamuro2012} is used to
reduce the FMOS data. We run the FIBRE-pac in a standard 
manner. One difference is that we first reduce the data per each set of
succeeding ``AB'' sequence, and then all of the data are
combined after the flux calibration while weighting each frame based 
on the flux observed through FMOS fibers. Note that the calibration
includes the correction for fiber flux loss which is estimated
from a calibration star.
The effects of differential refraction can be important at high
airmasses and with relatively small fibers (1.2" diameter). However,
the fiber positioner for FMOS is designed to mitigate these effects
\citep{Akiyama2008}. All of the observations were carried out at
airmass less than 2.0 and we re-position the fibers every 30 minutes
during our observations to account for the different zenith distances.  
Furthermore, we make sure that the stellar continuum spectrum is well
calibrated and there is no discrepancy in the flux between the spectra
separated into the $J$-long and $H$-long coverage for the stars used in
the flux calibration. 

For extended sources, additional aperture correction
is required to correct for flux loss. \citet{Yabe2012} have shown
that the covering fraction by the 1.2'' FMOS fiber is 0.45 for typical
galaxies at $z\sim1.4$. We use the same aperture correction in our
galaxies. Note that the covering fraction is not strongly dependent on
the size of the point spread function (PSF), if the galaxies are more
extended than 1.2''. As mentioned above, the flux calibration by
FIBRE-pac corrects for the flux loss for point-sources. The 1.2''
aperture includes $\sim$90\% of the total flux for the point sources.
Here we assume that the FWHM of the PSF is 0.85'' and a Gaussian
profile for the PSF, since the seeing in the FMOS observations are 
$\sim$0.8--0.9" in $H$-band.   
That is, we already corrected for the flux loss by $\sim$10\%. Thus,
we made aperture corrections of a factor of 2 (=0.9/0.45) for
both the $J$- and $H$-band spectra.   

Figure \ref{fig:individual_spectrum} shows the $J$- and $H$-band spectra
for an individual galaxy that has the largest observed [O\emissiontype{II}]
emission-line flux and all seven nebular emission lines
(H$\beta$, [O\emissiontype{III}]$\lambda\lambda$4959,5007, H$\alpha$, 
[N\emissiontype{II}]$\lambda$6584, and
[S\emissiontype{II}]$\lambda\lambda$6716,6731) are clearly detected.

\vspace{1em}
\subsubsection{Line detection}
\label{sec:Gaussian_fit}

\begin{figure}
 \begin{center}
  \includegraphics[width=8cm]{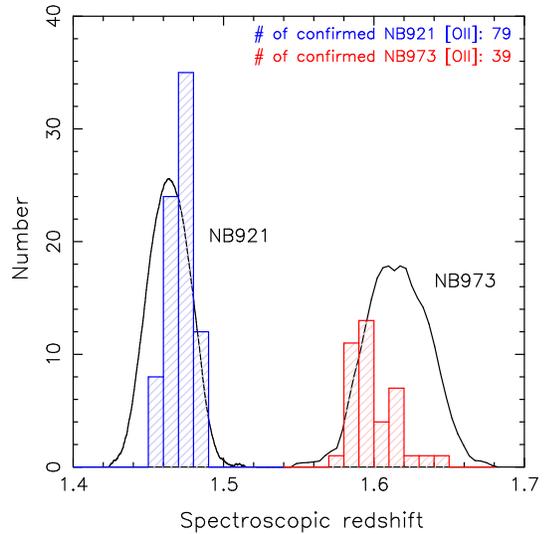} 
  \vspace{-1.0cm}
 \end{center}
\caption{
The distribution of redshifts confirmed by FMOS spectroscopy for the
118 [O\emissiontype{II}] emitters in the SDF. The blue histogram shows
the redshift distribution of 79 NB921 [O\emissiontype{II}] emitters,
while the red histogram shows that of 39 NB973 [O\emissiontype{II}]
emitters. The response curves of NB921 and NB973 narrow-band filters
are also shown, where the horizontal axis corresponds to the redshift from
which [O\emissiontype{II}] emission is shifted into the narrow-band
filter. 
}
\label{fig:redshift_distribution}
\end{figure}

We fit Gaussian profiles to the spectra to measure the redshift and
the flux of individual emission lines, and estimate the
signal-to-noise (S/N) ratio. 
The Gaussian fitting is conducted separately in the $J$ and $H$
spectra. In each band, major nebular emission lines are simultaneously
fitted; H$\beta$ and [O\emissiontype{III}]$\lambda\lambda$4959,5007 for
$J$-band, and H$\alpha$, [N\emissiontype{II}]$\lambda$$\lambda$6548,6584,
and [S\emissiontype{II}]$\lambda\lambda$6716,6731 for $H$-band. Note that
the width of the Gaussians fits is fixed to the same value of 
[O\emissiontype{III}]$\lambda5007$ or H$\alpha$. In addition, the flux
ratios of the [O\emissiontype{III}] and [N\emissiontype{II}] doublets
are fixed to one-third. Each emission line is well fitted
with single Gaussian profile with a velocity up to $\sim$100--200 km
s$^{-1}$ in $\sigma$, i.e., no broad component in the line, suggesting
that contamination from Type-1 AGN is negligible.  
Errors on the emission-line fluxes are estimated using the standard
deviation of 500 measurements of the flux, where in each measurement a
set of Gaussians is fitted to the spectrum to which the error with
normal distribution with 1$\sigma$ spectral noise is added. We regard
an emission line with S/N greater than 3 as a detection, and visually
inspect it for confirmation.  

We succeed in spectroscopically confirming 118
[O\emissiontype{II}] emitters at $z\sim1.5$ by detecting some emission
lines at more than 3$\sigma$, among which we have detected H$\alpha$ 
([O\emissiontype{III}]$\lambda$5007) for 115 (45) [O\emissiontype{II}] emitters. 
We also have detected H$\beta$, [N\emissiontype{II}]$\lambda$6584, and
[S\emissiontype{II}]$\lambda\lambda$6716,6731 in 13, 16,
and 6 galaxies, respectively.
The success rate of line detection is more than 60\% for galaxies with
observed [O\emissiontype{II}] flux larger than $\sim5\times10^{-17}$ 
erg s$^{-1}$ cm$^{-2}$ (Figure \ref{fig:target}). The
[O\emissiontype{III}] and H$\alpha$ emissions are the strongest lines
in $J$- and $H$-band spectra, respectively. Even if only a single emission
line is detected in the spectrum, it can be identified as being
[O\emissiontype{III}] or H$\alpha$, because of the additional detection of
[O\emissiontype{II}]$\lambda$3727 from narrow-band imaging. Thus, we
can investigate the luminosities of all major nebular emission
lines in the rest-frame optical for confirmed galaxies; 
i.e., [O\emissiontype{II}], H$\beta$, [O\emissiontype{III}],
H$\alpha$, [N\emissiontype{II}], and [S\emissiontype{II}].  
The emission-line fluxes measured for the 118 
confirmed galaxies are provided in Table \ref{tab:individual_obj}.

Spectroscopic redshifts are computed by taking an average of the
detected lines. Figure \ref{fig:redshift_distribution} shows the
redshift distribution for the 118 galaxies. We note that there is a
discrepancy between the peak of the redshift distribution of the
[O\emissiontype{II}] emitters and the response function of the
narrow-band filters.  The $H$-long observations do not cover
wavelengths $\lesssim$1.607$\mu$m in our
observations, and also the sensitivity declines sharply at the
wavelengths $\gtrsim$1.72$\mu$m. Thus, the H$\alpha$ emission
line is hard to detect at $z\lesssim$1.45 and
$z\gtrsim$1.62. This is a likely cause for the discrepancy between
the redshift distribution and the response of the narrow-band filters.  
Given an emission-line flux measured with narrow-band imaging,
galaxies located further away from the redshift corresponding to the
peak of the response curve will have larger intrinsic
line fluxes. 
That is, the redshift distribution shows that H$\alpha$ lines in
galaxies with larger intrinsic [O\emissiontype{II}] flux are likely to
be detected. This also suggests that it is important to correct
the emission-line flux for the filter response function using
spectroscopic redshift to measure the line flux accurately, even if
the observed flux is measured with narrow-band imaging.  

A fraction of observed galaxies has no emission line detected, which
may be due to several reasons. First, as described above, we are
likely to miss H$\alpha$ emission lines at $z\lesssim1.45$ or
$z\gtrsim1.62$.  
Second, the H$\alpha$ line strength is fainter than 
the detection limit. This would be the case 
for targets with observed [O\emissiontype{II}] flux lower than
$\sim5\times10^{-17}$ erg s$^{-1}$ cm$^{-2}$. Indeed,
since we regard an emission line at more than 3$\sigma$ as a detection,
we find some observed galaxies with marginal detection that are not
in excess of 3$\sigma$. 
Third, the flux of the line is reduced by the masking of OH
lines. We note that this is the case for the non-detection of
H$\alpha$ or [O\emissiontype{III}]$\lambda5007$ for some galaxies.  
Finally, it is possible that targeted galaxies are interlopers. However,
there are only two instances in NB921 [O\emissiontype{II}] emitters
where the [O\emissiontype{II}]$\lambda\lambda3726,3729$ doublet is
resolved and detected at $z=3.37$ and 3.72.
Thus, we conclude that there is low contamination rate in our samples
of [O\emissiontype{II}] emission-line galaxies at $z\sim1.5$.

\vspace{1em}
\subsubsection{Stellar mass estimates for [O\emissiontype{II}] emitters}
\label{sec:SED_fitting}

Multi-wavelength data from UV to mid-infrared are available in
the SDF. These data allow us to estimate the stellar mass for
individual galaxies by fitting the spectral energy distribution 
(SED) by population synthesis models. \citet{Ly2012}
performed the fitting to the near-UV-to-near-IR SED for the
[O\emissiontype{II}] emitters in this study. Thus, we briefly
summarize the procedure of their SED fitting. 

We used \citet{BC03} population synthesis models, and assumed the
\citet{Chabrier2003} initial mass function (IMF), exponentially
declining star formation histories with $\tau$=0.01, 0.1, 1.0, or 10 Gyr,
and solar metallicity. 
Note that the \citet{Chabrier2003} IMF is adopted throughout this paper,
although \citet{Salpeter1955} IMF is adopted in \citet{Ly2012}. 
The assumption in metallicity is justified given gas-phase metallicity
measurements of near solar values, as we will demonstrate later (\S
\ref{sec:N2}). We also note that the differences in stellar synthesis
models are minimal for different metal abundances.
The reddening curve of \citet{Calzetti2000} is assumed and dust
extinction of $A_V$=0.0--3.0 mag is considered.  

As shown in Figure \ref{fig:target}, the [O\emissiontype{II}] emitters
with emission line detected by FMOS have stellar masses of
$3\times10^8$ \MO\ to $2\times10^{11}$ \MO. With a range of three orders of 
magnitude in stellar mass, we are able to study both dwarf and massive
star-forming galaxies at $z\sim1.5$.

\begin{figure*}
 \begin{center}
  \includegraphics[width=15cm]{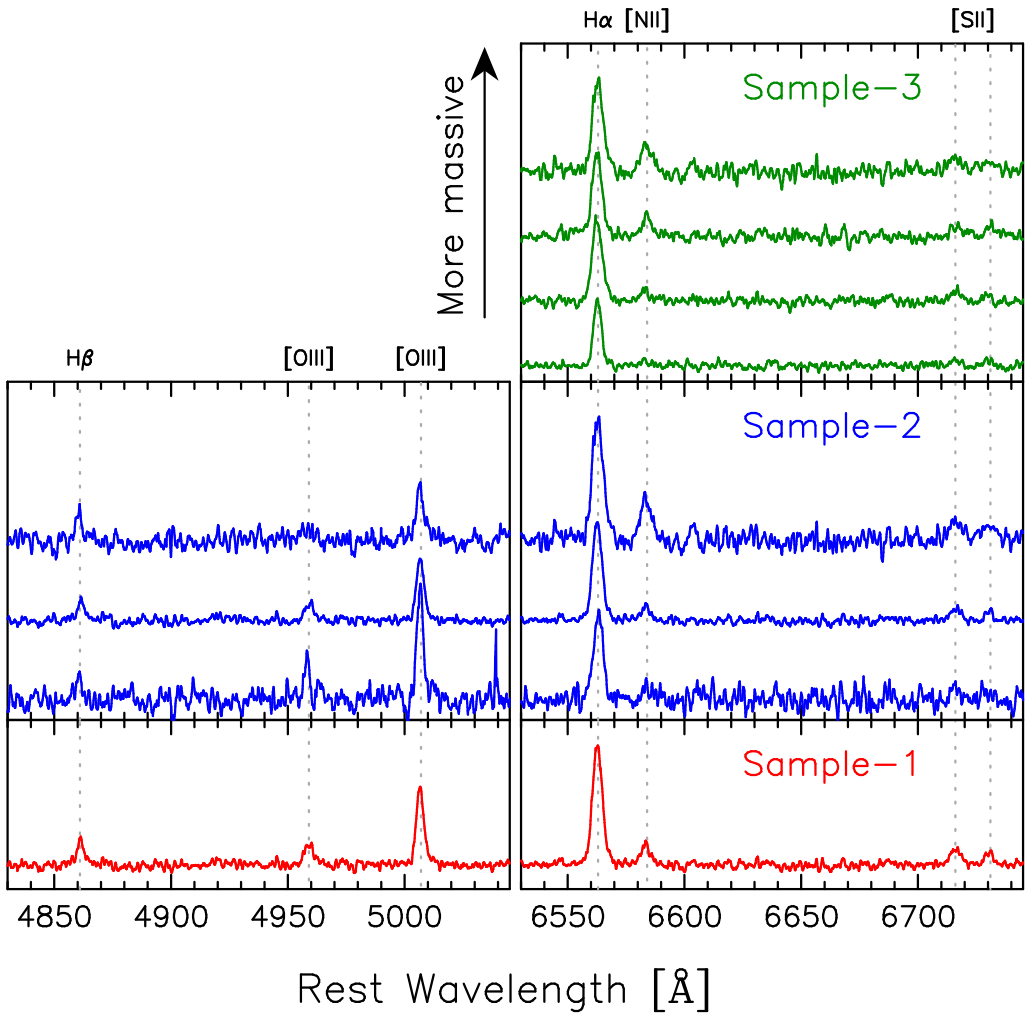} 
  \vspace{-1.5cm}
 \end{center}
\caption{
The stacked spectra for the subsamples summarized in Table
\ref{tab:stack_subsample}. The sample-1, -2, and -3 are shown in
bottom, middle, and top panels, respectively. In each panel, the
spectra are shifted vertically for illustration purposes and the
dotted lines show the wavelength of individual emission lines. 
The average stellar mass, $\langle\log(M_\star/M_\odot)\rangle$, of
the subsamples increases upward by 1.5 (1.1) dex in the sample-2 (3),
respectively. 
As galaxies are more massive, the [O\emissiontype{III}]$\lambda$5007
([N\emissiontype{II}]$\lambda$6584) line is weaker (stronger) relative
to H$\alpha$. 
}
\label{fig:stackedspectra}
\end{figure*}

\vspace{1em}
\subsubsection{Stacking FMOS spectra}
\label{sec;stacking}

Since H$\beta$, [N\emissiontype{II}] and [S\emissiontype{II}] are
generally weak for typical $z\sim1.5$ galaxies,
only one target has all five major nebular lines detected
(Figure \ref{fig:individual_spectrum}). In many
cases, H$\alpha$ line is the only emission line detected,
as shown in Tables \ref{tab:individual_obj} and
\ref{tab:stack_subsample}. Therefore, stacking analyses are essential
for investigating the average physical state of ISM in star-forming
galaxies at $z\sim1.5$.  

Table \ref{tab:stack_subsample} summarizes subsamples for stacking.
Sample-1 contains all [O\emissiontype{II}] emitters for which $J$- and
$H$-band spectra are available. Sample-2 consists of three subsamples
divided by stellar mass. The three subsamples in sample-2 are also
selected from all [O\emissiontype{II}] emitters for which $J$- and
$H$-band spectra are available. Sample-3 consists of four
subsamples divided by stellar mass, including all [O\emissiontype{II}]
emitters with $H$-band spectra, irrespective of $J$-band data. Sample-3 is
used for the analyses involving only H$\alpha$, [N\emissiontype{II}],
and [S\emissiontype{II}]. 
The subsamples are defined so that the fraction of the detected emission
line is as similar as possible, which avoids a large contribution of
a biased number of strong emission lines to the stacked spectra.      

To make the stacked spectra, we transform the observed spectra
into rest-frame based on the spectroscopic redshift. 
Then we compute the weighted mean of the spectra based on the noise at
each wavelength with 3$\sigma$ clipping. All stacked spectra, shown in
Figure \ref{fig:stackedspectra}, detect five nebular emission lines,
H$\beta$, [O\emissiontype{III}], H$\alpha$, [N\emissiontype{II}], and
[S\emissiontype{II}]. The line ratios in the individual stacked
spectra are shown in Table \ref{tab:lineratio}.

\subsection{The SDSS data}

In order to understand the physical state of ISM in star-forming
galaxies at $z\sim1.5$, comparisons with local galaxies are
useful. The local spectroscopic catalogs are extracted from the
MPA-JHU release for the Sloan Digital Sky Survey (SDSS) Data Release 7
(DR7) \footnote{\url{http://www.mpa-garching.mpg.de/SDSS/DR7/}}
\citep{Kauffmann2003b,Brinchmann2004,Salim2007,Abazajian2009}.
We use 86,238 objects that have S/N $\geq3$ for [O\emissiontype{II}],
H$\beta$, [O\emissiontype{III}], H$\alpha$, [N\emissiontype{II}], and 
[S\emissiontype{II}] with stellar mass larger than $10^8$\MO at
redshifts of 0.04--0.1. These objects are   
distinguished as star-forming galaxies, AGNs, or galaxies with
non-negligible contribution from AGNs (hereafter composite objects) on
the Baldwin-Phillips-Terlevich (``BPT'') diagram
([O\emissiontype{III}]/H$\beta$ vs. [N\emissiontype{II}]/H$\alpha$;
\citet{BPT1981}). 
The boundary defined by \citet{Kewley2001} is used to discriminate
star-forming galaxies and composite objects from AGNs, and then the
boundary defined by \citet{Kauffmann2003} is used to distinguish
between star-forming galaxies and composite objects (see also \S
\ref{sec:BPT}).    
The number of star-forming galaxies, composite objects, and AGNs are
67764, 10764, and 7710, respectively.

\section{Physical state of ISM in galaxies at $z\sim1.5$}
\label{sec:physical_state}

Six major emission lines in rest-frame optical are available for
star-forming galaxies at $z\sim1.5$ in this study, allowing us to
investigate their physical state of ISM in detail. We rely on stacked 
spectra to study the average properties, since most individual spectra
are not deep enough to detect all the nebular emission lines at
S/N$>$3. However, we also investigate individual measurements where
available. 

We first study the electron density of H\emissiontype{II} regions in
$z\sim1.5$ star-forming galaxies. Electron density has an impact on
line strengths.
The Balmer lines (e.g., H$\alpha$ and H$\beta$)
are emitted from recombination of ionized hydrogen gas, while
[O\emissiontype{III}] and [O\emissiontype{II}] are emitted by
collisional excitation to forbidden transition states followed by
radiative de-excitation.  
We then correct the emission lines for dust extinction. Correction for dust
extinction in the nebular emission is important for reliable
measurement of intrinsic ratios of nebular emission lines \citep[e.g.,][]{Ly2012b}.  
Next, we use several line ratios to examine possible contributions of AGN to the
emission lines (\S \ref{sec:BPT}), ionization parameter (\S
\ref{sec:ionization_parameter}), and metal abundance (\S
\ref{sec:metallicity}).
Ionization level is closely related to the intensity of the radiation
field in the nebular gas, while metal abundance reflects various
galaxy evolution processes.  
Finally, we investigate the relation between H$\alpha$ and
[O\emissiontype{II}] (\S \ref{sec:HaOII}), both of which are widely
used as indicators of star formation in galaxies, 
and the relation between stellar mass and SFR estimated from H$\alpha$ (\S \ref{sec:MS}). 
The H$\alpha$ luminosity directly reflects the number of ionizing
photons from massive young stars, while the [O\emissiontype{II}]
strength depends in several ways on the physical state of ISM, such as oxygen
abundance, ionization parameter, electron temperature, and electron
density. However, the use of [O\emissiontype{II}] has grown in
investigating star formation activity in high-$z$ galaxies, since
H$\alpha$ redshifts into NIR (MIR) wavebands for galaxies at
$z\gtrsim0.5$ ($z\gtrsim2.5$).

\subsection{Electron density}
\label{sec:electron_density}

The intensity ratios of the [O\emissiontype{II}]$\lambda\lambda$3726,3729
and [S\emissiontype{II}]$\lambda\lambda$6716,6731 doublets are sensitive
to the electron density of ISM. Note that these intensity ratios are
independent of dust extinction. Although we have the
[O\emissiontype{II}] luminosity for our full sample, narrow-band
imaging does not resolve the [O\emissiontype{II}] doublet. Instead, we
investigate the intensity ratio of the [S\emissiontype{II}]
doublet using the stacked spectra, and individual spectra for
bright galaxies. 

\begin{figure}
 \begin{center}
  \includegraphics[width=8cm]{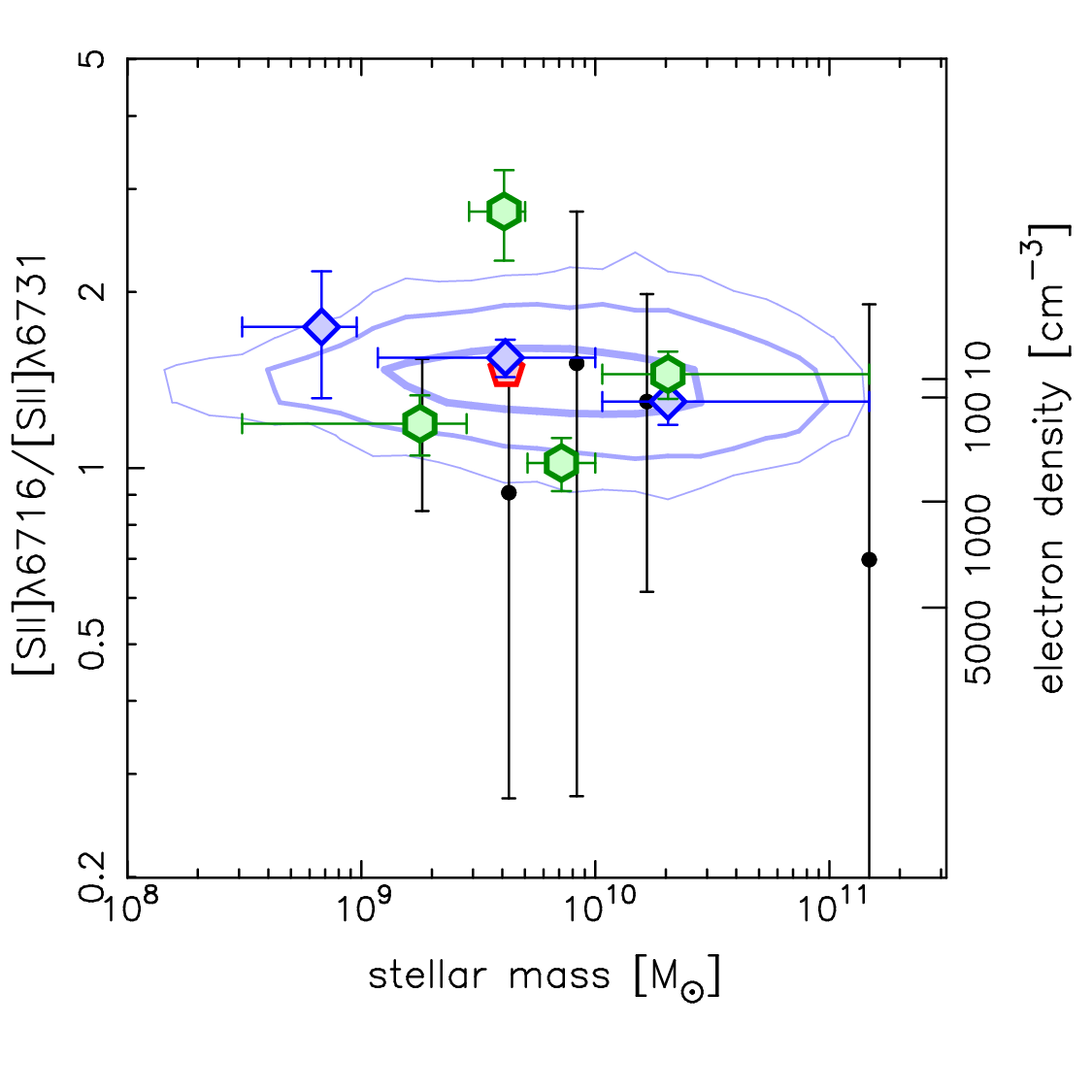} 
  \includegraphics[width=8cm]{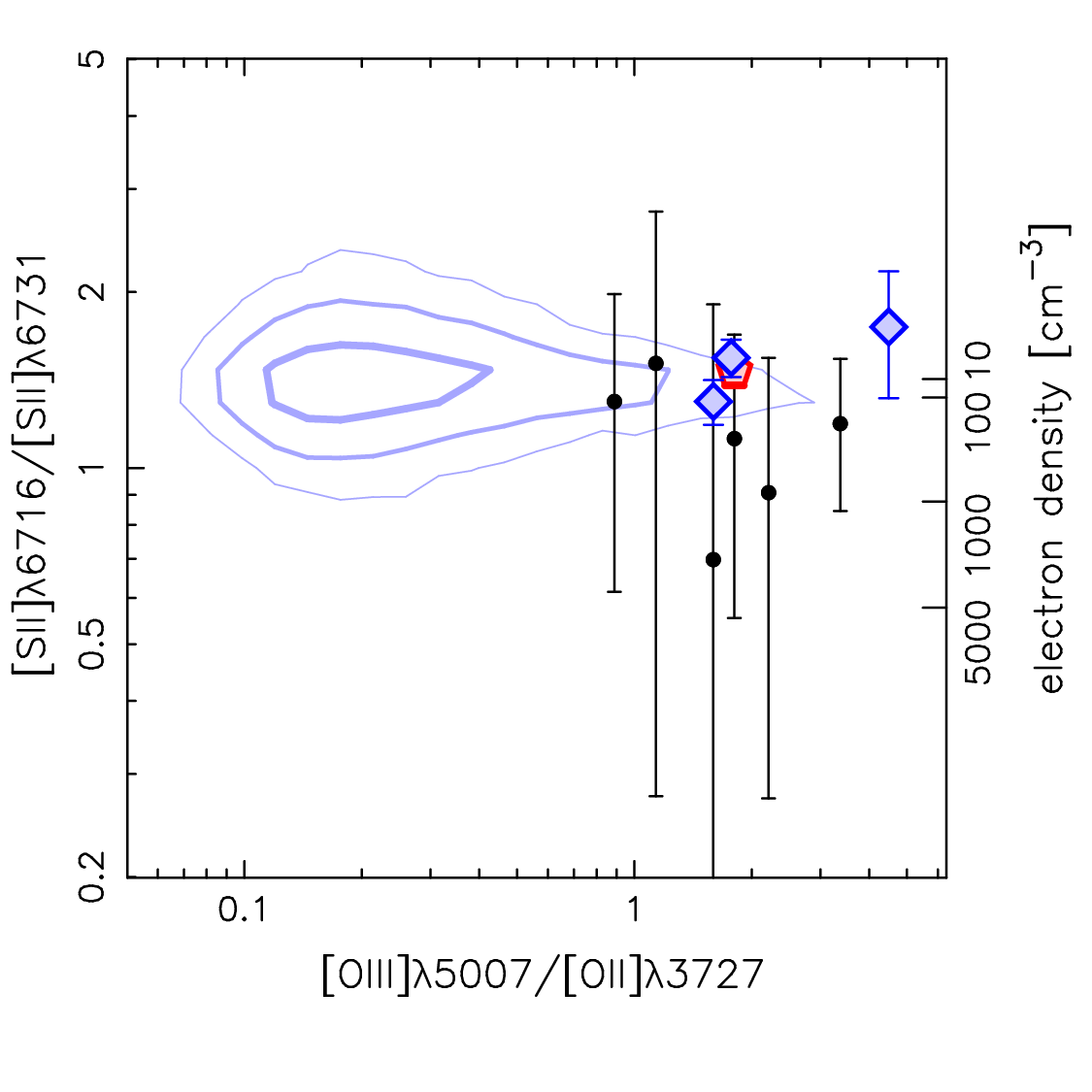} 
 \end{center}
 \vspace{-1.0cm}
\caption{
The line ratio of
[S\emissiontype{II}]$\lambda6716$/[S\emissiontype{II}]$\lambda6731$
as a function of stellar mass (top) and
[O\emissiontype{III}]/[O\emissiontype{II}] (bottom). 
The right axis shows the electron density $n_e$ corresponding to the line
ratio of [S\emissiontype{II}] doublet at $T_e=10^4$ K. 
The black filled circles show the ratio of the individual galaxies
with the doublet detected at $>3\sigma$. Red pentagon, blue diamonds,
and green hexagons show the average ratios derived from the stacked
spectra of sample-1, sample-2 and sample-3, respectively.
The contours show the distribution including 68\%, 95\%, and 99\% of
star-forming galaxies in the local Universe, where the thicker contour
includes the lower percentage of the population.  
}
\label{fig:ne}
\end{figure}

We find that the galaxies at $z\sim1.5$ have electron
densities ($n_e$) of $\sim10^2$ cm$^{-3}$. 
Figure \ref{fig:ne} shows that there is no clear dependence of
[S\emissiontype{II}]$\lambda$6716/[S\emissiontype{II}]$\lambda$6731 on
stellar mass of the galaxy, and suggests that electron density in
$z\sim1.5$ star-forming galaxies is consistent with typical estimates
for local galaxies, i.e., $n_e\sim$10--10$^2$ cm$^{-3}$
\citep{Osterbrock1989,Shirazi2014}.  
The electron density shown in the right axis of Figure \ref{fig:ne} is
estimated from the intensity ratio of the [S\emissiontype{II}] doublet
under the assumption of $T_e=10^4$ K, using the IRAF task {\it temden} 
\citep{Shaw1994}.
We also find that
[S\emissiontype{II}]$\lambda$6716/[S\emissiontype{II}]$\lambda$6731 is  
not dependent on [O\emissiontype{III}]/[O\emissiontype{II}],
suggesting no dependence on the ionization state of ISM (Figure \ref{fig:ne}). 
\citet{Masters2014} measure an electron density of
$n_e\simeq$100--400 cm$^{-3}$ for composite spectra of galaxies
at $\langle z \rangle=1.85$. \citet{Rigby2011} found an electron
density of $n_e=252^{+30}_{-28}$ cm$^{-3}$ from 
the ratio of the [O\emissiontype{II}] doublet in a $z=1.70$ lensed
star-forming galaxy.
The electron density of our [O\emissiontype{II}] emitters at
$z\sim1.5$ is consistent with those of galaxies at similar redshifts,
and with little evolution as a function of redshift. 

Some studies have found higher electron densities of $\sim10^3$
cm$^{-3}$ from [S\emissiontype{II}] doublet measurements for lensed
galaxies at $z\sim2$
\citep{Hainline2009,Bian2010}. \citet{Hainline2009} derive electron
densities of 320--1600 and 1270--2540 cm$^{-3}$ for the ``Cosmic 
Horseshoe'' ($z=2.38$) and the ``Clone'' ($z=2.00$),
respectively. \citet{Bian2010} measured electron densities of
1029$^{+3333}_{-669}$ and 1166$^{+7020}_{-855}$ cm$^{-3}$ for two
components of J0900+2234, a $z=2.03$ star-forming galaxy. 
We find that two individual galaxies have [S\emissiontype{II}] line
ratios showing larger electron density of 2.3$\times10^3$ or 8.9$\times10^2$
cm$^{-3}$ at $10^4$ K. 
The uncertainties in our individual measurements are large,
but our stacks suggest that such galaxies with denser gas are 
probably not in the majority.

\subsection{Dust extinction}
\label{sec:Balmer_decrement}

\begin{figure}
 \begin{center}
  \includegraphics[width=8cm]{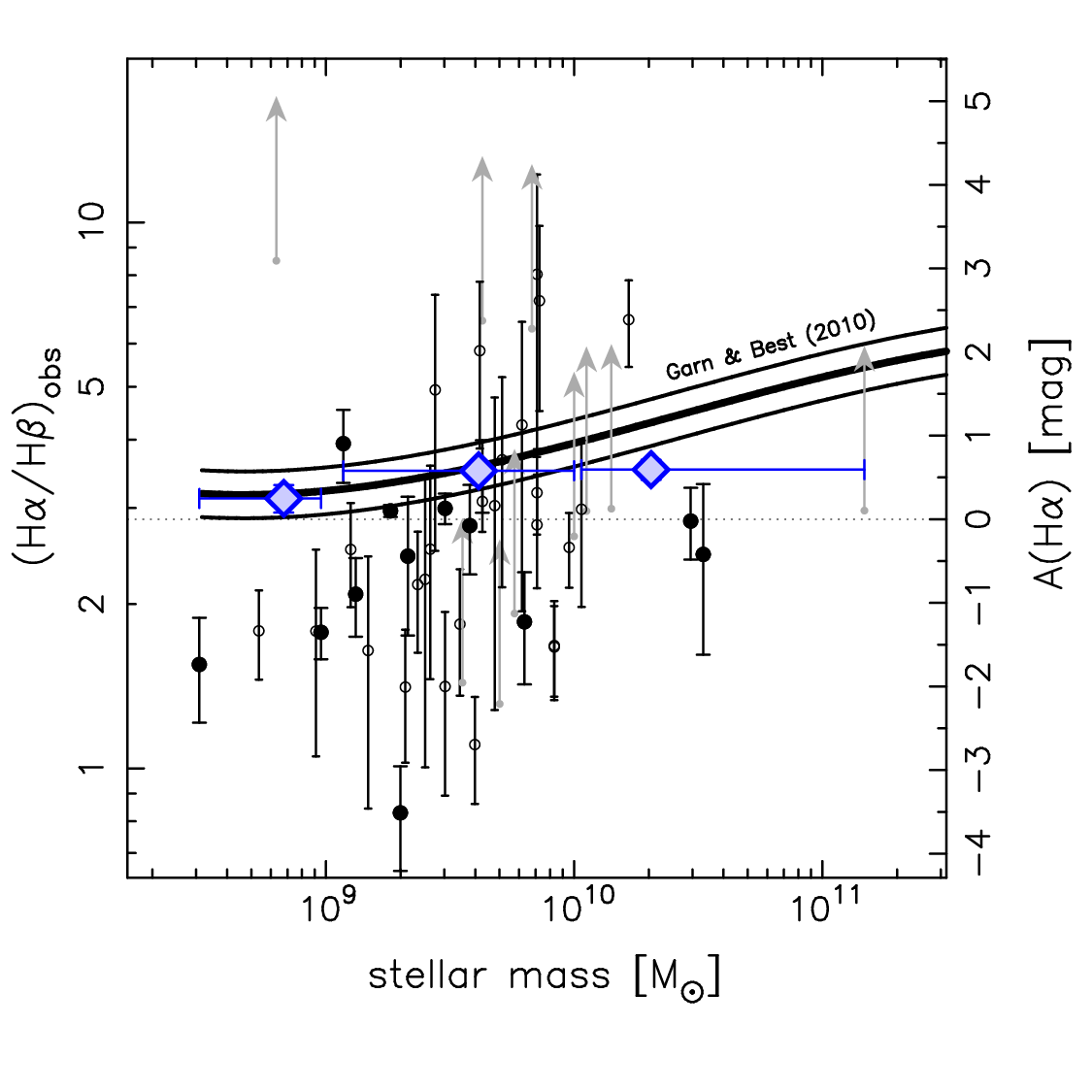} 
  \vspace{-1.0cm}
 \end{center}
\caption{
The observed ratio of H$\alpha$ to H$\beta$ (i.e., Balmer decrement),
as a function of stellar mass. The right axis shows extinction at
6563\AA\ corresponding to the line ratio of H$\alpha$ to H$\beta$, where
the reddening curve of \citet{Calzetti2000} is assumed.   
The black filled circles show the individual measurements with the two
lines detected at $>3\sigma$, while blue squares show the average
ratios derived from the stacked spectra (sample-2). 
The dotted line shows the intrinsic unreddened Case B value of
H$\alpha$/H$\beta$=2.86.
The stacked spectrum with all galaxies indicates an average line ratio of
$\langle$H$\alpha$/H$\beta$$\rangle$=3.57, resulting in
$\langle$A(H$\alpha$)$\rangle$=0.63. The open circles show the
galaxies with H$\beta$ detected at 2--3$\sigma$, while the arrows show
lower limits on the ratio using the 2$\sigma$ upper limit of
H$\beta$ for galaxies without H$\beta$ detections.
The solid curves show the relation between A(H$\alpha$) and stellar
mass in the local Universe and its $\pm1\sigma$ distribution given by
\citet{GarnBest2010}.  
}
\label{fig:HaHb}
\end{figure}

The difference between the observed and intrinsic ratios of H$\alpha$
to H$\beta$, i.e., Balmer decrement, can be used to estimate the
attenuation of nebular emission by dust. Assuming an electron temperature of
$T_e=10^4$ K and electron density of $n_e=10^2$ cm$^{-3}$ for Case B
recombination, the intrinsic ratio of H$\alpha$/H$\beta$ is 2.86
\citep{Osterbrock1989}. In converting the value of Balmer decrement
into extinction at 6563\AA, we assume the reddening curve of
\citet{Calzetti2000}. 

Figure \ref{fig:HaHb} shows the observed ratio of H$\alpha$/H$\beta$
as a function of stellar mass. Note that no correction is made
for the absorption of stellar continuum at the H$\alpha$ and H$\beta$
wavelengths. The stellar continuum is not detected in
individual FMOS spectra, and it is hard to estimate the influence of
the absorption of stellar continuum on the Balmer lines reliably with
only our data. Thus, we assume that the absorption is weak enough for
the correction not to be required, if any \citep[see also, e.g.,][]{Steidel2014}.   

For individual measurements, about half of them have observed
H$\alpha$/H$\beta$ ratios that are consistent with
the unreddened Case B value of 2.86 within the uncertainties. However,
many galaxies have line ratios of H$\alpha$/H$\beta<$2.86. The
reason for the lower line ratio is unclear, however, it
is possible that the error is larger than what is shown in the figure,
due to the uncertainty in positioning of individual fibers allocated
to the galaxies. While assuming a higher electron temperature
$T_e\gtrsim10^4$ K can slightly reduce the intrinsic
H$\alpha$/H$\beta$, the ratio only decreases to 2.74 for 
$T_e=2\times10^4$ K and $n_e=10^2$ cm$^{-3}$ \citep{Osterbrock1989}.
The results of individual dust extinction measurements with both H$\alpha$ and
H$\beta$ available can be biased towards relatively low
H$\alpha$/H$\beta$ ratios of $\lesssim2.86$, since H$\beta$ is
easier to detect with less dust extinction or
intrinsically large H$\beta$ luminosity. 

On the other hand, the stacked spectra appear to follow the relation
between A(H$\alpha$) and stellar mass given by \citet{GarnBest2010},
although the extinction in the highest mass bin is lower than the
relation. 
As suggested by the dual narrow-band survey simultaneously targeting
H$\alpha$ and [O\emissiontype{II}] for star-forming galaxies at
$z\approx1.47$ \citep{Sobral2012}, we also confirm that the relation
of \citet{GarnBest2010} is applicable to most star-forming
galaxies at $z\sim1.5$ using the Balmer decrement \citep[see also][]{Dominguez2013,Price2014}. 
With the stacked spectrum of sample-1, we find that our
[O\emissiontype{II}] emitters have an average
dust extinction of $\langle$A(H$\alpha$)$\rangle=0.63$.    

In this paper, unless otherwise specified, we adopt the following
procedure to correct for dust extinction of nebular emission. First,
we assume the dust extinction curve given by \citet{Calzetti2000}. 
For galaxies with both H$\alpha$ and H$\beta$ detected at S/N$>$3,
the dust extinction is estimated from Balmer decrement. 
However, if (H$\alpha$/H$\beta$)$_{\rm obs}<$2.86, we
assume zero dust extinction. For galaxies without an observed
H$\alpha$/H$\beta$ ratio, the \citet{GarnBest2010} relation between
A(H$\alpha$) and stellar mass is used.

\subsection{BPT diagnostics diagram}
\label{sec:BPT}

\begin{figure}
 \begin{center}
  \includegraphics[width=8cm]{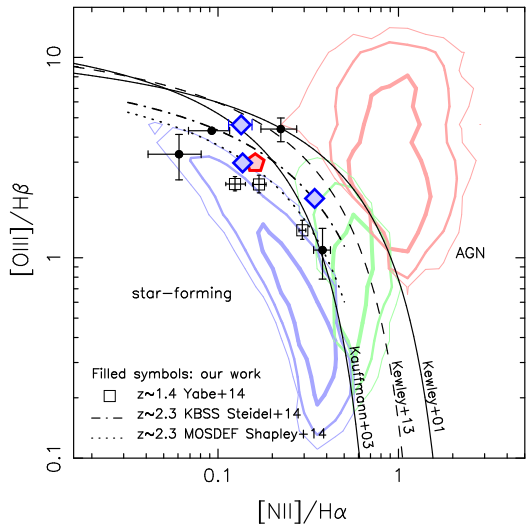} 
 \end{center}
 \vspace{-1cm}
\caption{
The BPT diagram showing the [O\emissiontype{III}]/H$\beta$  ratios versus the
[N\emissiontype{II}]/H$\alpha$ ratio. The black circles show the  
individual galaxies with all of the four lines detected, while red
pentagon and blue diamonds show the average ratios derived from the
stacked spectra of sample-1 and sample-2, respectively. 
Open squares show the line ratios for the stacked spectra of mass-selected
star-forming galaxies at $z\sim1.4$ \citep{Yabe2014}. The dashed-dotted
and dotted lines show the relations fitted to $z\sim2.3$ star-forming
galaxies in the Keck Baryonic Structure Survey
\citep[KBSS;][]{Steidel2014} and MOSFIRE Deep Evolution Field
\citep[MOSDEF;][]{Shapley2014} survey, respectively.
The contours show the distribution including 68\%, 95\%, and 99\% of
star-forming galaxies (blue), composite objects (green), and AGNs (red)
in the local Universe, respectively, where the thicker contour
includes the lower percentage of the population.    
The solid lines are the boundaries defined by
\citet{Kauffmann2003} and \citet{Kewley2001} to distinguish the
populations at low redshift. The dashed line is the upper limit of the distribution of
star-forming galaxies at $z=1.5$ that \citet{Kewley2013b}
proposed. 
}
\label{fig:BPT}
\end{figure}

The BPT diagram that compares
[O\emissiontype{III}]$\lambda$5007/H$\beta$ against
[N\emissiontype{II}]$\lambda$6584/H$\alpha$ \citep{BPT1981} is
frequently used to classify galaxies and understand whether massive
young stars, shocks, or AGN power the ionization of the gas.
We illustrate in Figure \ref{fig:BPT} the average BPT line ratios, as well as,
measurements for individual galaxies if all four lines are
detected at S/N$>$3. Here, we apply the boundary defined by
\citet{Kewley2001} to identify AGNs, and use \citet{Kauffmann2003}'s
boundary to distinguish star-forming galaxies
from composite objects and AGNs. The figure shows that none of our
galaxies are located in the AGN region. This result is consistent with
the low velocity dispersion of the emission lines, which
indicates that our galaxies are not Type-1 AGN (\S \ref{sec:Gaussian_fit}).
Since the FMOS fibers measure integrated light, there is still a
possibility that the galaxies host an obscured or weak AGN. In such
a case, the AGN would not be expected to contribute significantly 
to the emission lines we observe. 

Several recent studies have reported an offset on the BPT diagram for 
star-forming galaxies at $z\approx1.5$--2.5,
compared with that of local galaxies 
\citep[e.g.,][]{Erb2006a,Rigby2011,Dominguez2013,Newman2014,Shapley2014,Steidel2014,Yabe2014,Zahid2014}.
Figure \ref{fig:BPT} shows that our [O\emissiontype{II}] emitters at
$z\sim1.5$ also have an offset on the BPT diagram,
which is consistent with previous studies.
From a photoionization modeling perspective, \citet{Kewley2013b}
derive a redshift-dependent boundary to separate purely star-forming
galaxies from galaxies hosting an AGN. The revised boundary classifies
our [O\emissiontype{II}] emitters at $z\sim1.5$ as star-forming
galaxies (Figure \ref{fig:BPT}).   

\citet{Kewley2013} discusses the influence of the ionization parameter,
the electron density, and the shape of UV spectrum on the BPT line
ratios. \citet{Kewley2013} finds that a higher electron density and a
steeper UV spectral slope can result in both BPT line ratios being
larger. A higher ionization parameter leads to an upward shift in
[O\emissiontype{III}]/H$\beta$ in the range of
[N\emissiontype{II}]/H$\alpha$ that our $z\sim1.5$
[O\emissiontype{II}] emitters span. 

It is an interesting question whether the offset in the BPT diagram is in
[N\emissiontype{II}]/H$\alpha$ or [O\emissiontype{III}]/H$\beta$.  
We attribute most of this ``BPT offset'' to a higher
[O\emissiontype{III}]/H$\beta$ ratio at $z\sim1.5$ than in the 
local Universe. This is because at a given stellar mass, our
[O\emissiontype{II}] emitters have [N\emissiontype{II}]/H$\alpha$
ratios that are similar to or slightly smaller than that of local
galaxies \citep[\S \ref{sec:metallicity} and see also ][]{Stott2013,Yabe2014,Zahid2014}.    
This implies that an upward shift of the
[O\emissiontype{III}]/H$\beta$ ratio is necessarily required to match
the distribution of the local galaxies with that of the
[O\emissiontype{II}] emitters at $z\sim1.5$, at a given stellar mass.   
Furthermore, the absence of strong cosmic evolution in the typical
line ratio of the [S\emissiontype{II}] doublet suggests
that electron density is not responsible for the BPT offset,
as shown in \S \ref{sec:electron_density}. 
One possible explanation is a larger ionization
parameter.  In the next section, we
investigate the ionization parameter for the [O\emissiontype{II}]
emitters at $z\sim1.5$, and discuss whether this interpretation is
correct.

\begin{figure*}
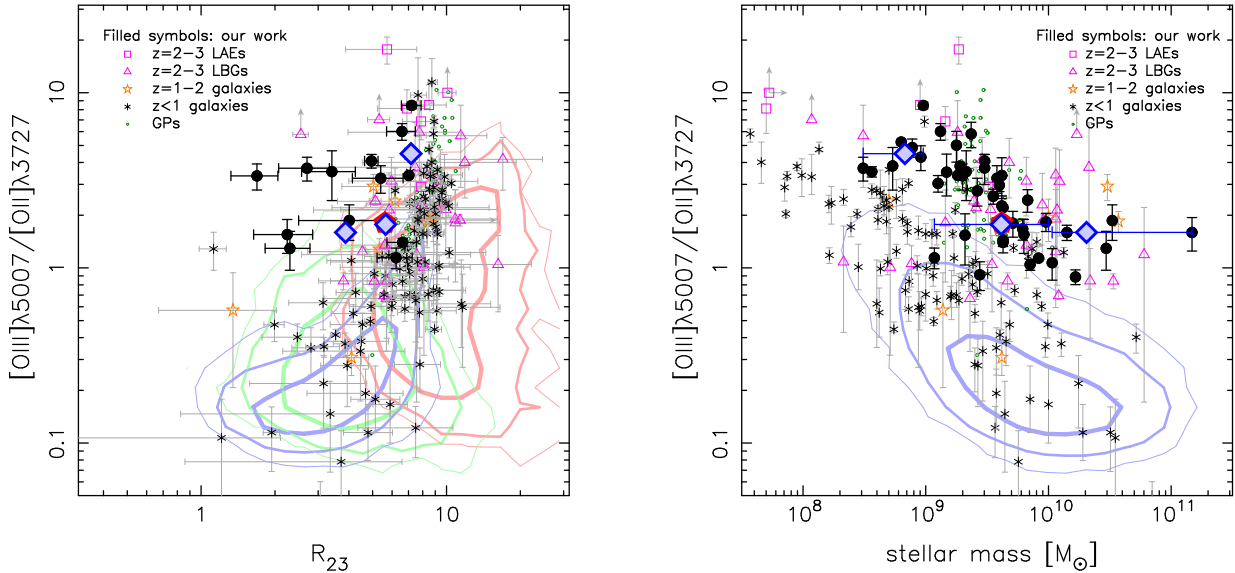

 \begin{center}
  \includegraphics[width=8.7cm]{./fig8a.eps} 
  \includegraphics[width=8.7cm]{./fig8b.eps} 
 \end{center}
 \vspace{-1cm}
\caption{
The line ratio of [O\emissiontype{III}]/[O\emissiontype{II}], which is
sensitive to ionization parameter, as a function of $R_{23}$ (left)
and stellar mass (right). 
The black circles show the ratio of the individual galaxies with all
of the required lines detected, while red pentagon and blue diamonds
show the average ratios derived from the stacked spectra of sample-1
and sample-2, respectively. 
Different populations up to $z\sim3$ are also shown;
LAEs at $z=$2--3 \citep[open squares:][]{Fosbury2003,Erb2010,Christensen2012b,Nakajima2013},
LBGs at $z=$2--3 \citep[open triangles:][]{Pettini2001,Maiolino2008,Mannucci2009,Richard2011,Wuyts2012,Belli2013}, 
star-forming galaxies at $z=$1--2 \citep[open stars:][]{Queyrel2009,Rigby2011,Christensen2012b}, 
star-forming galaxies at $z<1$ \citep[asterisks:][]{Savaglio2005,Maier2005,Henry2013,Ly2014,Ly2014b}, 
and GPs \citep[open circles:][]{Cardamone2009}.
The contours show the distribution including 68\%, 95\%, and 99\% of
star-forming galaxies (blue), composite objects (green), and AGNs (red)
in the local Universe, respectively, where the thicker 
contour includes the lower percentage of the population.   
}
\label{fig:ionization_parameter}
\end{figure*}

\subsection{Ionization parameter}
\label{sec:ionization_parameter}

It is known that some galaxies at low redshifts, so-called 
``Green Peas'' \citep[GPs;][]{Cardamone2009}, have extremely high
[O\emissiontype{III}] equivalent width. These green peas are rare, low-mass
systems with low metallicity 
\citep{Cardamone2009,Amorin2010}. 
Recent studies have found high-$z$ galaxies that possess
strong [O\emissiontype{III}] emission
\citep{Ly2007,Ly2014,Ly2014b,vandelWel2011,Nakajima2013,Nakajima2014,Richardson2013,Atek2014,Amorin2014,Holden2014,Shirazi2014}.    
The ratios of nebular emission lines of star-forming galaxies
at $z$=1--3 on the BPT diagram imply that typical galaxies at
high redshifts tend to have stronger [O\emissiontype{III}] emission,
than local star-forming galaxies (Figure \ref{fig:BPT}).   

The ionization parameter is defined as 
\begin{eqnarray}
q\equiv\frac{Q_{H^0}}{4\pi R_s^2n_H},
\label{eq:q}
\end{eqnarray}
where $Q_{H^0}$ is the number of ionizing photons above the Lyman
limit per unit time, $R_s$ is the Str$\ddot{\rm o}$mgren radius, and
$n_H$ is the hydrogen density. 
The ratio of
[O\emissiontype{III}]$\lambda$5007 to [O\emissiontype{II}]$\lambda$3727
is sensitive to the ionization parameter of the ISM \citep{McGaugh1991,Kobulnicky2004,Nakajima2014}. 
The effective temperature of ionizing sources also affects the
[O\emissiontype{III}]/[O\emissiontype{II}] ratio, since a harder UV photon
can ionize O$^+$ \citep{PerezMontero2005}.  

The left panel of Figure \ref{fig:ionization_parameter} shows the
ratio of [O\emissiontype{III}]/[O\emissiontype{II}] against the $R_{23}$
index, ([O\emissiontype{II}]$\lambda$3727+[O\emissiontype{III}]$\lambda\lambda$4959,5007)/H$\beta$, for
galaxies with the required line detected. In the local Universe,
$\sim0.1$\% of SDSS galaxies have 
[O\emissiontype{III}]/[O\emissiontype{II}] $>1.0$, and a ratio of
[O\emissiontype{III}]/[O\emissiontype{II}] $\gtrsim3.0$ is seldom
seen. Compared with local star-forming galaxies,
almost all of our [O\emissiontype{II}] emitters have higher
[O\emissiontype{III}]/[O\emissiontype{II}] ratios of $\gtrsim 1$ at all
$R_{23}$ values. 
The [O\emissiontype{III}]/[O\emissiontype{II}] ratios of our
[O\emissiontype{II}] emitters at $z\sim1.5$ are larger than those of
typical galaxies at $z<1$, and consistent with those of galaxies at $z=1$--2
\citep{Queyrel2009,Rigby2011,Christensen2012b}, as well as, Lyman
$\alpha$ emitters (LAEs) and Lyman break galaxies (LBGs) at
$z\sim$2--3 \citep{Richardson2013,Amorin2014,Nakajima2014}.  

The right panel of Figure \ref{fig:ionization_parameter}
shows the dependence of
[O\emissiontype{III}]/[O\emissiontype{II}] on stellar mass.
The contours and asterisks in the figure suggest that there is a similar
dependence of [O\emissiontype{III}]/[O\emissiontype{II}] on stellar
mass for galaxies at $z=0.04$--0.1 and for star-forming galaxies at
$z<1$ \citep{Henry2013,Ly2014,Ly2014b}. 
At a given stellar mass, the
[O\emissiontype{III}]/[O\emissiontype{II}] ratios of our
[O\emissiontype{II}] emitters are larger than those of typical
galaxies at $z<1$, and consistent with those of most galaxies at
$z=1$--2 as well as LBGs at $z\sim$2--3. 
Massive star-forming galaxies with 
stellar mass of $\sim10^{10}$\MO\ at $z\sim1.5$ appear to have
[O\emissiontype{III}]/[O\emissiontype{II}] ratios that are similar to
less massive star-forming galaxies with stellar mass of
$\sim10^{9}$\MO\ in the local Universe.    

We measure the ionization parameter, $q$ in units of cm s$^{-1}$, from the
line intensity ratios of $R_{23}$ and $O_{32}$
([O\emissiontype{III}]$\lambda\lambda4959,5007$/[O\emissiontype{II}]$\lambda3727$),  
using the equations given in \citet{Kobulnicky2004} where the
photoionization model of \citet{Kewley2002} is used. 
[O\emissiontype{II}] emitters at $z\sim1.5$ have 
$\langle$$\log(q/{\rm cm\ s^{-1}})$$\rangle$=8.1, 
while star-forming galaxies in the local Universe have  
$\langle\log(q/{\rm cm\ s^{-1}})\rangle\sim7.3$ \citep[e.g.,][]{Dopita2006,Nakajima2014}.
Here, we assume that [O\emissiontype{II}] emitters follow the
high-metallicity branch of the relation between metallicity and $R_{23}$
(see \S\ref{sec:metallicity}).
Even if the low-metallicity branch is assumed, average ionization
parameter of [O\emissiontype{II}] emitters is 
$\langle$$\log(q/{\rm cm\ s^{-1}})$$\rangle$=7.7.  
Thus, we conclude that 
[O\emissiontype{III}]/[O\emissiontype{II}] ratios of
[O\emissiontype{II}] emitters at $z\sim1.5$ may be increased by higher 
ionization parameters. We discuss possible mechanisms causing the high
ionization parameter in high-$z$ galaxies in \S \ref{sec:discussion}.

\subsection{Gas-phase metallicity}
\label{sec:metallicity}

Gas-phase metallicity derived from electron temperature measurements
are considered to be the most reliable.  
This is because metals aid in the cooling of gas by optically-thin
radiation, and thus electron temperature is sensitive to the gas-phase
metallicity.   
However, such measurements are difficult to obtain, because the
emission line allowing us to estimate the
electron temperature is too weak to be detected for individual
galaxies\footnote{Some studies have succeeded in measuring or
providing a strong constraint on the electron temperature 
\citep[e.g.,][]{Kakazu2007,Hu2009,Yuan2009,Rigby2011,Christensen2012a,Ly2014,Ly2014b,Amorin2014b}}.  
Therefore, several ``strong-line'' emission-line diagnostics, which 
are calibrated with local galaxies, are widely used to
derive gas-phase metallicity of high-$z$ galaxies. Here, we will
assume that these diagnostics are applicable for high-$z$ galaxies.

However, we must be cautious in interpreting high-$z$ metallicity studies. 
In particular, the measured ionization parameter appears to evolve with
redshift (\S \ref{sec:ionization_parameter}). Furthermore, recent
studies have examined the nitrogen-to-oxygen abundance (N/O), and find
enhancement relative to local galaxies with the same oxygen abundance
\citep{Teplitz2000,Amorin2010,Masters2014,Shapley2014}.   

\begin{figure}
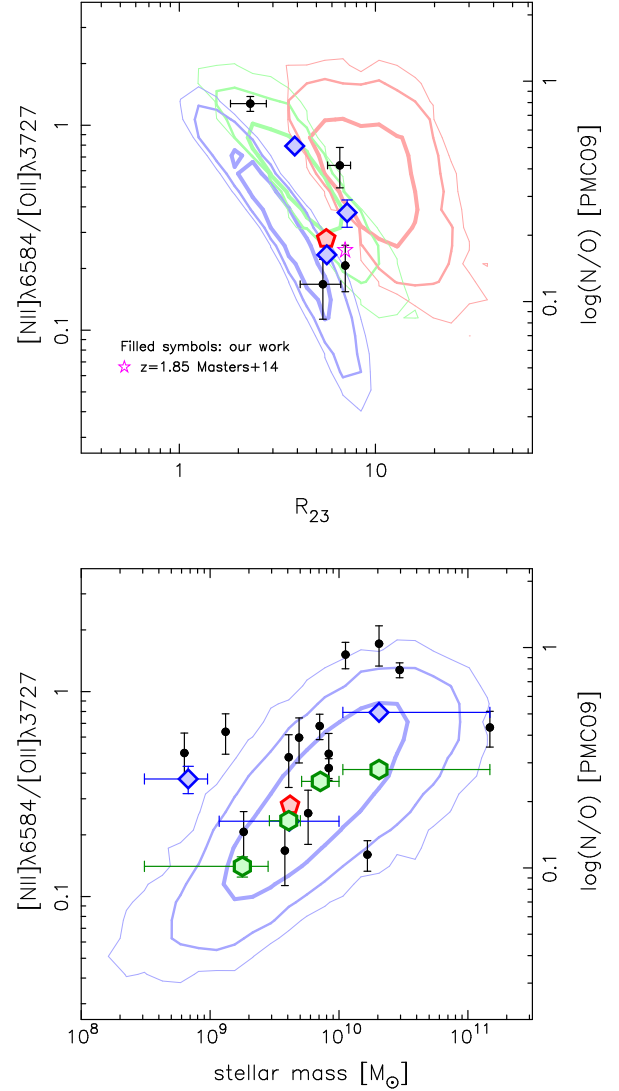

 \begin{center}
  \includegraphics[width=8cm]{./fig9a.eps}\\[-5mm] 
  \includegraphics[width=8cm]{./fig9b.eps} 
 \end{center}
 \vspace{-1cm}
\caption{
The line ratio of
[N\emissiontype{II}]$\lambda6584$/[O\emissiontype{II}]$\lambda3727$
as a function of $R_{23}$ (top) and stellar mass (bottom). The line ratio is correlated 
with the nitrogen-to-oxygen ratio (N/O), while $R_{23}$ is an indicator
of oxygen abundance, O/H. 
The black circles show the ratio of the individual galaxies with all
of the required lines detected, while the red pentagon, blue diamonds, and
green hexagons show the average ratios derived from the stacked
spectra of sample-1, sample-2 and sample-3, respectively. 
The star shows the average ratio for
composite spectrum at $\langle z \rangle=1.85$ \citep{Masters2014}.
The right axis shows the N/O ratio based on calibration by
\citet{PerezMontero2009}. 
The contours show the distribution including 68\%, 95\%, and 99\% of
star-forming galaxies (blue), composite objects (green), and AGNs (red)
in the local Universe, respectively, where the thicker 
contour includes the lower percentage of the population.   
}
\label{fig:N/O}
\end{figure}

\begin{figure}
 \begin{center}
  \includegraphics[width=8cm]{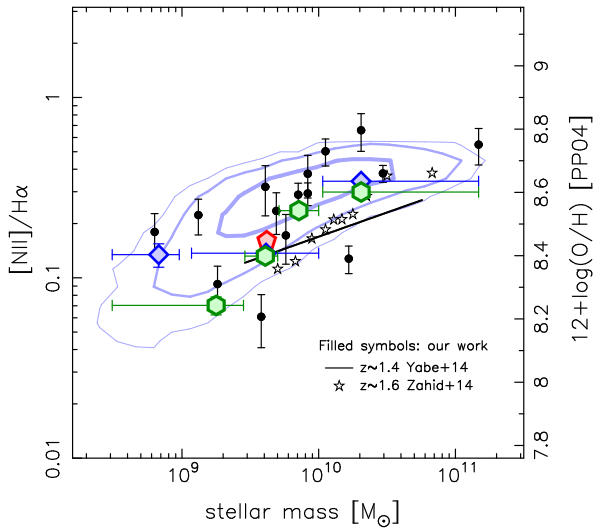} 
  \vspace{-1cm}
 \end{center}
\caption{
The line ratio of [N\emissiontype{II}]/H$\alpha$ as a function of
stellar mass. The vertical axis on the right shows the oxygen
abundance corresponding to the N2 index, using the calibration given by
\citet{Pettini2004}.   
The black filled circles show the ratio of the individual galaxies
with the doublet detected at $>3\sigma$. The red pentagon, blue diamonds,
and green hexagons show the average ratios derived from the stacked
spectra of sample-1, sample-2 and sample-3, respectively.
The solid line and stars show the relations given by \citet{Yabe2014}
and \citet{Zahid2014} for star-forming galaxies at $z\sim1.4$ and
$z\sim1.6$, respectively. The contours show the distribution including
68\%, 95\%, and 99\% of star-forming galaxies in the local Universe,
where the thicker contour includes the lower percentage of the
population.   
}
\label{fig:N2}
\end{figure}

\vspace{1em}
\subsubsection{N/O abundance ratio}

The [N\emissiontype{II}]$\lambda6584$/[O\emissiontype{II}]$\lambda3727$ ratio
is correlated with the nitrogen-to-oxygen abundance ratio in
H\emissiontype{II} regions \citep{PerezMontero2005,PerezMontero2009}. 
Since the ionizing potential energies for nitrogen and
oxygen are similar, the line intensity ratio is insensitive to
ionization conditions \citep{Kewley2002}. 
The relation between the N/O abundance ratio estimated from the [N\emissiontype{II}]
and [O\emissiontype{II}] electron temperatures and
[N\emissiontype{II}]$\lambda6584$/[O\emissiontype{II}]$\lambda3727$ is
derived by \citet{PerezMontero2009}. 

Figure \ref{fig:N/O} shows 
[N\emissiontype{II}]$\lambda6584$/[O\emissiontype{II}]$\lambda3727$
as a function of $R_{23}$, that is, the nitrogen-to-oxygen (N/O) ratio
as a function of the oxygen abundance (O/H). The distribution of galaxies
in the local Universe is overlaid in this figure. At a given oxygen
abundance, the [O\emissiontype{II}] emitters at $z\sim1.5$ appear to
have higher [N\emissiontype{II}]/[O\emissiontype{II}] relative to local
star-forming galaxies,
and the line ratios of the [O\emissiontype{II}] emitters at $z\sim1.5$
are similar to those of local composite objects.  
While local AGNs (red contours) have stronger [N\emissiontype{II}]
emission, compared with star-forming galaxies, the contribution of
an AGN in our [O\emissiontype{II}] emitters is negligible.  
We made certain that similar results are obtained even if
[N\emissiontype{II}]$\lambda6584$/[S\emissiontype{II}]$\lambda\lambda6716,6731$
is used as a indicator of the N/O abundance ratio
\citep{PerezMontero2009}, instead of [N\emissiontype{II}]$\lambda6584$/[O\emissiontype{II}]$\lambda3727$.

Given the result that galaxies at $z\sim1.5$ have higher
ionization parameter than local galaxies, the $R_{23}$ value
does not necessarily measure metallicity evolution. Higher
ionization parameter leads to lower metallicity at a given $R_{23}$
\citep{Kewley2002}, suggesting that oxygen abundance of high-$z$
galaxies is lower than that of local galaxies. Nevertheless, the
[O\emissiontype{II}] emitters at $z\sim1.5$ are likely to have higher
N/O abundance ratio than local star-forming galaxies. The result that
we find is similar to recent findings of \citet{Amorin2010} for GPs
and \citet{Masters2014} for star-forming galaxies at
$z\sim1.85$. \citet{Teplitz2000} also found high nitrogen abundance
for MS 1512-cB58, which is a gravitationally lensed starburst galaxy
at $z=2.72$.    

Figure \ref{fig:N/O} also shows that there is no significant
difference in dependence of N/O abundance ratio on stellar mass
between the [O\emissiontype{II}] emitters at $z\sim1.5$ and local
star-forming galaxies. This suggests that at a given stellar mass, the
[O\emissiontype{II}] emitters have similar N/O abundance ratio to
local star-forming galaxies, which is the same as the recent finding
of \citet{Amorin2010} for GPs.     
On the other hand, \citet{PerezMontero2013} found evidence for redshift
evolution in the relation between N/O abundance ratio and stellar mass
for $z\lesssim0.4$. 
In particular, they argue that galaxies at $z=$0.2--0.4 are offset in
N/O abundances from local galaxies by -0.14$\pm$0.31 dex. While this
is in disagreement with our result for [O\emissiontype{II}] emitters
at $z\sim1.5$, we note that the selection functions of our sample and
\citet{PerezMontero2013} are different, and it is not clear if these
differences may affect N/O abundance measurements.

\citet{PerezMontero2009} found that it is likely that some
star-forming galaxies are classified as composite objects due to high
[N\emissiontype{II}]/H$\alpha$ ratio resulting from high N/O abundance
ratio. The SDSS subsample of star-forming galaxies may be biased
toward star-forming galaxies with low N/O abundance ratio, which can
lead us to underestimate the N/O abundance ratio of local star-forming
galaxies. This suggests that it is possible that at a given
metallicity the difference in N/O abundance ratio between the
[O\emissiontype{II}] emitters at $z\sim1.5$ and local star-forming
galaxies is smaller than what is shown in Figure \ref{fig:N/O}. 
Also, this may result in the slight evolution of the relation between
N/O abundance ratio and stellar mass, as seen by \citet{PerezMontero2013}.
However, Figure \ref{fig:N/O} shows that the line ratios of
[N\emissiontype{II}]$\lambda6584$/[O\emissiontype{II}]$\lambda3727$  
are similar to those of local composite objects, although in the BPT
diagram (Figure \ref{fig:BPT}), most [O\emissiontype{II}] emitters at
$z\sim1.5$ are located in the star-forming region defined by \citet{Kauffmann2003}.
Thus, we cannot completely rule out the possibility that the
[O\emissiontype{II}] emitters at $z\sim1.5$ have higher N/O abundance
ratio at a given metallicity than local star-forming galaxies, while
having similar N/O abundance ratio at a given stellar mass.
The trends of N/O abundance ratio against oxygen abundance and stellar
mass, as shown in Figure \ref{fig:N/O}, appear to be due to the
evolution of mass-metallicity relation between the
[O\emissiontype{II}] emitters at $z\sim1.5$ and local star-forming
galaxies.

\begin{figure*}
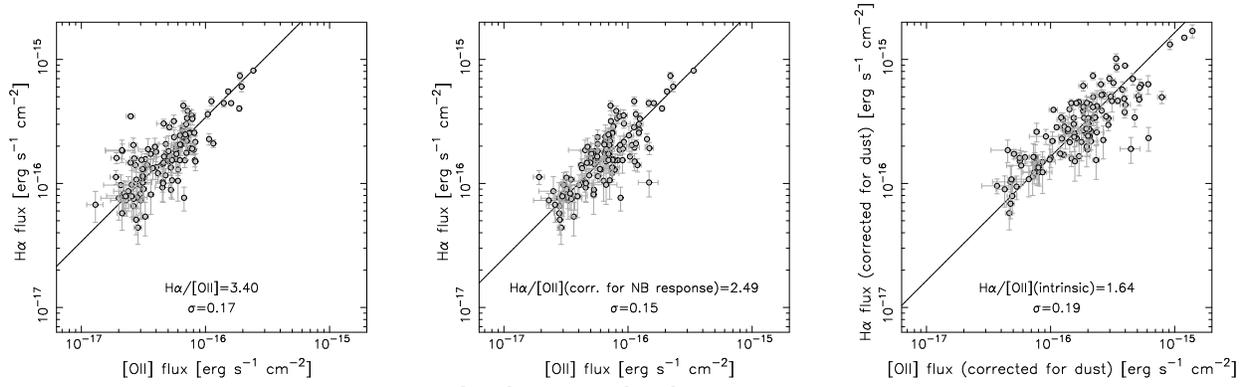

 \begin{center}
  \includegraphics[width=5.5cm]{./fig11a.eps} 
  \includegraphics[width=5.5cm]{./fig11b.eps} 
  \includegraphics[width=5.5cm]{./fig11c.eps} 
 \end{center}
\vspace{-1cm}
\caption{
The flux of H$\alpha$ is shown against [O\emissiontype{II}] flux for
[O\emissiontype{II}] emitters at $z\sim1.5$. In the left panel,
H$\alpha$ flux is not corrected for dust extinction (i.e., observed
flux), and [O\emissiontype{II}] flux is measured from narrow-band
imaging (i.e., not corrected for anything). The middle panel is the
same as the left one, but for [O\emissiontype{II}] flux being
corrected for the filter response function based on the spectroscopic
redshift. In the right panel, H$\alpha$ and [O\emissiontype{II}] are
both corrected for dust extinction (see the text in \S
\ref{sec:Balmer_decrement} for the details of the procedure of dust
correction).  
}
\label{fig:Halpha_oii}
\end{figure*}

\vspace{1em}
\subsubsection{Oxygen abundance from [N\emissiontype{II}]$\lambda6584$/H$\alpha$}
\label{sec:N2}

Figure \ref{fig:N2} shows the
[N\emissiontype{II}]$\lambda6584$/H$\alpha$ ratio as a function of
stellar mass. The N2 index,
$\log$([N\emissiontype{II}]$\lambda6584$/H$\alpha$), is widely used as
a proxy for oxygen abundance at high-$z$  
\citep[e.g.,][]{Denicolo2002,Pettini2004}.
Since different diagnostics can lead to different metallicity estimates
\citep{Kewley2008}, one of the merits of using the N2 index is ease
of a fair comparison between our result and previous studies  
\citep[e.g.,][]{Erb2006a,hayashi2009,Stott2013,Yabe2014,Zahid2014}. 
Figure \ref{fig:N2} shows that there is a dependence of the
[N\emissiontype{II}]/H$\alpha$ ratio on stellar mass for both
individual galaxies and the average from stacked spectra. The
[N\emissiontype{II}]/H$\alpha$ distribution for individual galaxies at
$z\sim1.5$ is similar to that of local galaxies. We note that
individual measurements are biased because of requiring
[N\emissiontype{II}] emission.   
Thus, the results from stacked spectra (green symbols in the figure) show
the proper, average relation for [O\emissiontype{II}] emitters at
$z\sim1.5$. Our [O\emissiontype{II}] emitters at $z\sim1.5$ show
steeper increase of the line ratio with the stellar mass, compared to
local galaxies. The difference in the
[N\emissiontype{II}]/H$\alpha$ ratio between the two populations is
0.1 dex at the highest stellar mass bin, while the difference is 0.4
dex at the smallest stellar mass bin. 

Higher ionization parameter leads to a lower N2 index at a given
metallicity \citep{Kewley2002}. With evidence that there is a redshift
evolution of the ionization parameter, the difference of the N2 index
between various galaxy populations at different redshifts does not
necessarily imply a metallicity difference.  
Since less massive [O\emissiontype{II}] emitters at 
$z\sim1.5$ tend to have higher ionization parameter, the metallicity
for less massive galaxies may be more underestimated. We also find
that the nitrogen-to-oxygen abundance ratio for [O\emissiontype{II}]
emitters at $z\sim1.5$ is possibly higher than that of local
galaxies, which can lead to overestimating the metallicity
\citep{PerezMontero2009,Morales-Luis2014}. Thus, all we can conclude from our data is
that there is a mass-metallicity relation that extends toward $M_\star\sim10^{9}$ \MO\ at $z\sim1.5$.

Photoionization modeling allows us to take into account the dependence of line ratios
on the ionization parameter in estimating the metallicity
\citep[e.g.,][]{McGaugh1991,Charlot2001,Kewley2002}. 
The $R_{23}$ index and the relationship between metallicity,
ionization parameter, and the line intensity ratios,
which is derived from photoionization models, enable estimates of
oxygen abundance without the concerns 
described above \citep{Kobulnicky2004}, although at a given $R_{23}$
value there are two metallicity solutions of low and high metallicities.
We will estimate oxygen abundance more reliably with $R_{23}$ index, and then
discuss the mass-metallicity relation and the correlation between
stellar mass, metallicity, and SFR for our $z\sim1.5$ [O\emissiontype{II}]
emitters  in a forthcoming paper.

\subsection{The ratio of H$\alpha$ to [O\emissiontype{II}]}
\label{sec:HaOII}

\citet{Hayashi2013} studied the relation between H$\alpha$ and
[O\emissiontype{II}] for galaxies at $z=1.47$, and concluded that the
luminosity of [O\emissiontype{II}] is well correlated with that of
H$\alpha$, and thus [O\emissiontype{II}] can be used to estimate the star
formation activity even at $z=1.47$ \citep[see also][]{Lee2012,Sobral2012}. 
However, this study is based on the stacking analysis of narrow-band
H$\alpha$ images for [O\emissiontype{II}] emitters at $z=1.47$.  
Therefore, it is worth investigating the relation between
H$\alpha$ and [O\emissiontype{II}] for 115 individual galaxies
at $z\sim1.5$. 

Figure \ref{fig:Halpha_oii} shows the H$\alpha$ flux as
a function of [O\emissiontype{II}] flux for individual galaxies. 
The observed fluxes of H$\alpha$ and [O\emissiontype{II}] are shown in
the left panel. There is a clear correlation between H$\alpha$ and
[O\emissiontype{II}] with a low dispersion in the
H$\alpha$/[O\emissiontype{II}] ratio of $\sigma=0.17$.    
However, as discussed in \S \ref{sec:Gaussian_fit}, the
[O\emissiontype{II}] observed fluxes require corrections for the response
function of narrow-band filter. The middle panel demonstrates the
importance of the correction, which ranges from 1.0 to 4.7,
depending on the redshift of the [O\emissiontype{II}] emitters (Figure
\ref{fig:redshift_distribution}).
It shows that the correlation between the H$\alpha$ and [O\emissiontype{II}]
observed fluxes is slightly tighter, $\sigma=0.15$. 
We also investigate the intrinsic ratio of H$\alpha$/[O\emissiontype{II}]
by correcting for dust attenuation (\S \ref{sec:Balmer_decrement}). 
The right panel shows again the tight correlation where the ratio is
$1.64\pm0.19$, which is consistent with the line 
ratio of local galaxies \citep{Kennicutt1998,Moustakas2006} and
with $z\sim2$ H$\alpha$-selected galaxies \citep{Lee2012}. 
Note that the luminosity range of H$\alpha$ and [O\emissiontype{II}]
covered by individual galaxy samples is different by more than an order
of magnitude between at $z\sim1.5$ and at $z\sim0$. Nevertheless, 
the tight correlation between H$\alpha$ and [O\emissiontype{II}] at
$z\sim1.5$ implies that the [O\emissiontype{II}] luminosity can be used to
estimate star formation activity in galaxies at $z\sim1.5$, similar to that
of local galaxies. This is consistent with the result of \citet{Ly2012}
who compared [O\emissiontype{II}] and UV SFRs in $z\sim1.5$
[O\emissiontype{II}] emitters.

\begin{figure}
 \begin{center}
  \includegraphics[width=8cm]{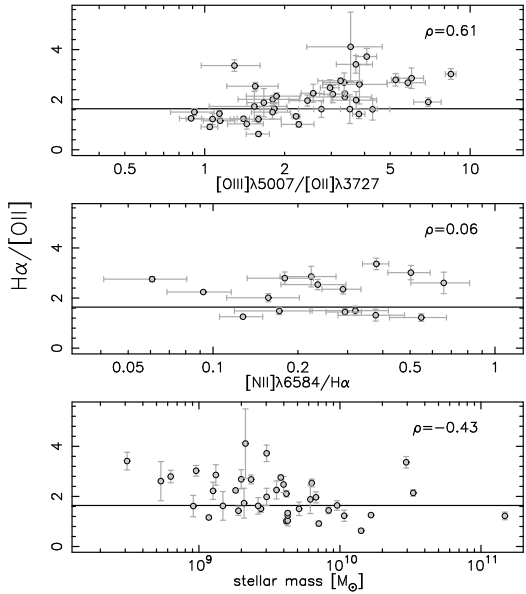} 
 \end{center}
\caption{
The dust-corrected H$\alpha$/[O\emissiontype{II}] as a function of
stellar mass (bottom), [N\emissiontype{II}]/H$\alpha$ (middle), and
[O\emissiontype{III}]/[O\emissiontype{II}] (top). 
The solid line shows the H$\alpha$/[O\emissiontype{II}] ratio of 1.64.
The number in the upper right corner of each panel shows Spearman's
rank correlation coefficient. The ratio of
[N\emissiontype{II}]/H$\alpha$ is a proxy of metallicity, while the
ratio of [O\emissiontype{III}]/[O\emissiontype{II}] is a proxy of
ionization parameter. The ratio of H$\alpha$/[O\emissiontype{II}]
shows a significant dependence on stellar mass and ionization parameter.  
}
\label{fig:DELHaoii}
\end{figure}

Even if the ratio of H$\alpha$/[O\emissiontype{II}] is corrected for
dust extinction, there is a non-negligible dispersion ($\sigma=0.19$;
Figure \ref{fig:Halpha_oii}). Since the H$\alpha$/[O\emissiontype{II}]
ratio can depend on the physical conditions in H\emissiontype{II}
regions 
\citep[e.g.,][]{Moustakas2006,Weiner2007,Gilbank2010,Maier2014b},
we compare the ratio to various properties. 
Figure \ref{fig:DELHaoii} shows that the
H$\alpha$/[O\emissiontype{II}] ratio depends on stellar mass and
the ionization parameter, [O\emissiontype{III}]/[O\emissiontype{II}]
(Spearman's rank correlation coefficient is $\rho=-0.43$ and
$\rho=0.61$, respectively), while there seems to be no strong
correlation between H$\alpha$/[O\emissiontype{II}] and
[N\emissiontype{II}]/H$\alpha$, i.e., gas metallicity
($\rho=0.06$). Thus, Figure \ref{fig:DELHaoii} suggests that less
massive galaxies or galaxies with higher ionization parameter tend to
have larger H$\alpha$/[O\emissiontype{II}]. 

Given the earlier result that less massive galaxies have larger ionization
parameter (\S \ref{sec:ionization_parameter}),  
the dependence of H$\alpha$/[O\emissiontype{II}] on both stellar mass
and [O\emissiontype{III}]/[O\emissiontype{II}] appears to be related
to the stellar mass--ionization parameter dependence.  
Since a higher ionization parameter results in a larger O$^{++}$ ionic
population relative to O$^+$, it is natural to expect a
weak [O\emissiontype{II}] emission for galaxies with high
ionization parameter. 
Indeed, we notice that at a given H$\alpha$ flux, galaxies with larger
[O\emissiontype{III}]/[O\emissiontype{II}] ratio tend to have lower  
[O\emissiontype{II}] flux. This suggests a dependence of the
H$\alpha$/[O\emissiontype{II}] ratio on the ionization parameter.  

The mass-metallicity relation may
cause the stellar-mass dependence of H$\alpha$/[O\emissiontype{II}].
However, the relation between metallicity and $R_{23}$ index
\citep[e.g.,][]{Pagel1979} implies that the emission from oxygen ions
([O\emissiontype{II}] and [O\emissiontype{III}]) becomes weaker against
Balmer emission (H$\beta$) with increasing metallicity. We here assume
that the galaxies follow the high-metallicity branch of the relation
between metallicity and $R_{23}$ index. If this is true, the
mass-metallicity relation should lead to a positive correlation
between H$\alpha$/[O\emissiontype{II}] and stellar mass, which is
contrary to what we observe. 

We conclude that although the intrinsic ratio of
H$\alpha$/[O\emissiontype{II}] for star-forming galaxies at $z\sim1.5$
is consistent with that of local galaxies, the ionization
parameter has an impact on the H$\alpha$/[O\emissiontype{II}] ratio in
H\emissiontype{II} regions.  

\begin{figure}
 \begin{center}
  \includegraphics[width=8cm]{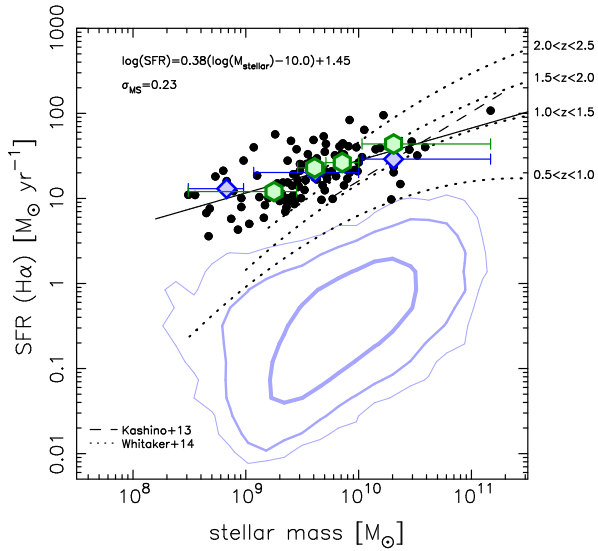} 
 \end{center}
\vspace{-10mm}
\caption{
SFR as a function of stellar mass. 
The SFRs are derived from H$\alpha$ luminosities using the
conversion of \citet{Kennicutt1998}, with corrections for dust
attenuation as discussed in Section \ref{sec:Balmer_decrement}.
The black filled circles show the [O\emissiontype{II}] emitters at
$z\sim1.5$. Blue diamonds, and green hexagons show the average SFRs
derived from the stacked spectra of sample-2 and sample-3, 
respectively. The solid line is derived from a linear fit to the
individual [O\emissiontype{II}] emitters; $\log({\rm
  SFR})=0.38\times\log(M\star/10^{10}\MO)+1.45$ with a dispersion of
0.23 dex. 
The contours show the distribution including 68\%, 95\%, and 99\% of
star-forming galaxies in the local Universe, where the thicker contour
includes the lower percentage of the population.  
The broken line shows the relation for star-forming galaxies at
$z\sim1.6$ obtained from Subaru/FMOS spectroscopy \citep{Kashino2013},
while the dotted lines are the relations for star-forming galaxies
at $z=$0.5--1.0, 1.0--1.5, 1.5--2.0 and 2.0--2.5, respectively
\citep{Whitaker2014}.  
}
\label{fig:MS}
\end{figure}

\subsection{Correlation between SFR and stellar mass}
\label{sec:MS}

It is well known that normal star-forming galaxies at each redshift
follow a  correlation between SFR and stellar mass
from $z\sim0$ up to $z\sim3$, which is 
sometimes called the ``main sequence of star formation'' or the
``star-forming sequence''
\citep[e.g.,][]{Daddi2007,Noeske2007,Salim2007,Peng2010,Whitaker2012,Whitaker2014}.  
We investigate the SFRs of the [O\emissiontype{II}] emitters at
$z\sim1.5$ as a function of stellar mass, where the SFRs are derived
from the dust-corrected H$\alpha$ luminosities using the
conversion given in \citet{Kennicutt1998}.     
Note that we adjust the conversion factor to that based on
\citet{Chabrier2003} IMF to estimate SFRs from H$\alpha$ luminosities.
Figure \ref{fig:MS} shows that there is a correlation between the SFRs
and stellar masses, suggesting that almost all of the
[O\emissiontype{II}] emitters are not extreme galaxies but normal
star-forming galaxies which follow the main sequence at their redshift. 
From fitting a linear function to the individual [O\emissiontype{II}]
emitters, the main sequence (MS) is expressed as
$\log({\rm SFR})=0.38\times\log(M_{\star}/10^{10}\MO)+1.45$ with the
dispersion of $\sigma_{\rm MS}=0.23$ dex.
Such a ``main sequence'' of star-forming galaxies at $z\sim1.5$ is
consistent with other studies
\citep[e.g.,][]{Kashino2013,Whitaker2014}. However, the slope of the
relation that we have obtained for the [O\emissiontype{II}] emitters is
shallower than other studies (Figure \ref{fig:MS}). This seems
due to the selection bias for the [O\emissiontype{II}] emitters
with H$\alpha$ emission detected by our NIR spectroscopy.  

The local star-forming galaxies are also plotted in Figure
\ref{fig:MS}. Although the [O\emissiontype{II}] emitters have been
demonstrated to be typical star-forming galaxies at $z\sim1.5$ 
\citep[see also][]{Ly2012}, the comparison with the local
galaxies indicates that at a given stellar mass, star-forming galaxies
at $z\sim1.5$ have SFRs a factor of $\sim$30 times larger than typical
local star-forming galaxies. We discuss the effect of the enhancement
of star-forming activity at higher redshifts on the ISM of high-$z$
star-forming galaxies in \S \ref{sec:cause_of_high-q}.

\section{Discussion}
\label{sec:discussion}

Unlike typical local galaxies, we have found that star-forming
galaxies at $z\sim1.5$ have [O\emissiontype{III}]/[O\emissiontype{II}]
$\gtrsim1$ and a large ionization parameter at all stellar masses. We
discuss the origin of the large ionization parameter for star-forming
galaxies at $z\sim1.5$, and then compare the line ratios for our
star-forming galaxies at $z\sim1.5$ against several NIR spectroscopic
studies of star-forming galaxies at $z\approx$ 1.5--2.5.

\begin{figure*}
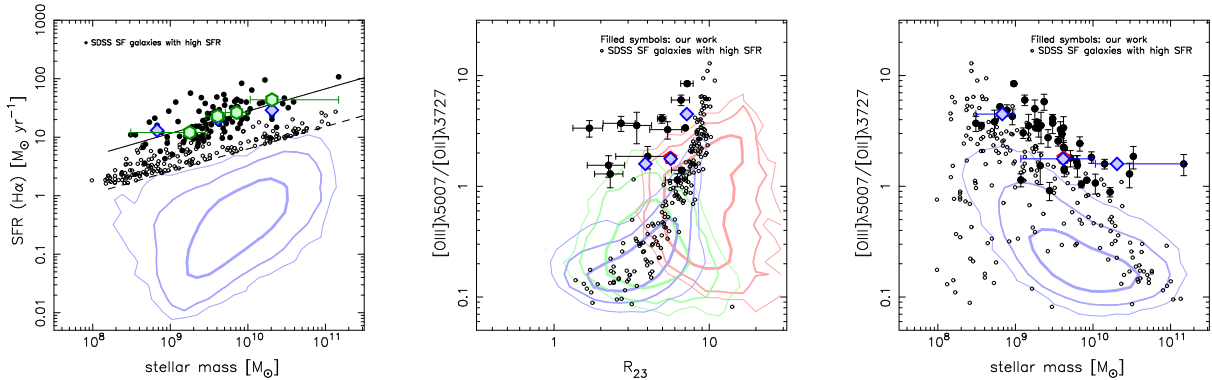

 \begin{center}
  \includegraphics[width=5.5cm]{./fig14a.eps} 
  \includegraphics[width=5.5cm]{./fig14b.eps} 
  \includegraphics[width=5.5cm]{./fig14c.eps} 
 \end{center}
\vspace{-5mm}
\caption{
The left panel is the same as Figure \ref{fig:MS} and the middle and right
panels are the same as Figures \ref{fig:ionization_parameter}, but for
the SDSS star-forming galaxies with high SFRs comparable to the
[O\emissiontype{II}] emitters at $z\sim1.5$  also plotted. To
compare the line ratios between star-forming galaxies with similar
SFRs at $z\sim0$ and $z\sim1.5$, SDSS star-forming galaxies meeting the
criterion (dashed line in the left panel),
$\log({\rm SFR})>0.38\times\log(M_{\star}/10^{10}\MO)+0.8$, 
are selected (open circles). 
}
\label{fig:comp_highSFR}
\end{figure*}

\subsection{The cause of large ionization parameter}
\label{sec:cause_of_high-q}

Several mechanisms have been considered to explain the high ionization state
\citep[e.g.,][]{PerezMontero2005,Kewley2013,Nakajima2014,PerezMontero2014}:
the shape of the ionizing radiation field which depends on an
effective temperature of ionizing sources and metallicity, the
geometry of nebular gas, and the hydrogen density.  
Since we find the electron density of H\emissiontype{II} regions in
galaxies at $z\sim1.5$ is similar to that of local galaxies (\S
\ref{sec:electron_density}), it is unlikely that there is a significant
difference in hydrogen density between star-forming galaxies at
$z\sim0$ and 1.5.  

First, we consider the geometry of nebular gas. Equation (\ref{eq:q})
states that a smaller Str$\ddot{\rm o}$mgren radius can result in
a higher ionization parameter. We begin with the simple assumption
that the nebular gas has a homogeneous distribution around young stars
in H\emissiontype{II} regions. In an ionization-bounded scenario, the
radius of H\emissiontype{II} regions is the Str$\ddot{\rm o}$mgren
sphere radius, which is determined by an equilibrium between photoionization and
recombination. In the density-bounded case, the radius is
determined by the size of gas clouds, which is smaller than the
Str$\ddot{\rm o}$mgren radius. 
Thus, at a given hydrogen density, density-bounded H\emissiontype{II}
regions are more able to explain high ionization parameters than
ionization-bounded regions.
We note that the actual distribution of nebular gas is likely to be inhomogeneous.
If the distribution of nebular gas within H\emissiontype{II} regions is
clumpy, then a density-bounded scenario is feasible.
However, a more detailed discussion on the geometry of nebular gas is
not possible with the existing seeing-limited data.

Another possible explanation for the high ionization parameter is due to
the radiation field. Of course, AGNs can result in a higher radiation
field, and shocks from galactic winds can trigger additional
excitation of the ionized gas \citep[e.g.,][]{Rich2011,Fogarty2012}.  
However, the emission-line ratios and line profiles suggest that an AGN
hard radiation field is negligible (Figures \ref{fig:stackedspectra}
and \ref{fig:BPT}).  
Since lower metallicity, which is found in star-forming galaxies
at $z\sim1.5$, can lead to a harder radiation field, metallicity may
play an important role in determining ionization parameter. Indeed,
there is a correlation between ionization parameter and metallicity
\citep{PerezMontero2014}. However, the left panel of Figure
\ref{fig:ionization_parameter} indicates that even at a given
$R_{23}$ index (oxygen abundance), star-forming galaxies at $z\sim1.5$ have larger
ionization parameter than galaxies at lower redshifts.

Both SFR and specific SFR (SFR/M$_{\star}$) at a
given stellar mass increase with redshift up to $z\sim$ 2--3, and
less massive galaxies have higher specific SFRs.
\citep[e.g.,][and see also Figure \ref{fig:MS}]{Whitaker2012,Speagle2014,Ilbert2014,Tasca2014}.   
Higher star-formation activity corresponds to an increasing number
of ionizing photons, which could result in a larger ionization
parameter at $z\sim1.5$ if the geometry and density of gas do not
change.

To investigate how the high star-forming activity affects the
ionization state of ISM, we compare the [O\emissiontype{II}] emitters
with local star-forming galaxies with comparably large SFRs.
They are selected as
galaxies with $\log({\rm SFR})>0.38\times\log(M_{\star}/10^{10}\MO)+0.8$
(Figure \ref{fig:comp_highSFR}). Only a very small fraction of
local galaxies meet this criterion.  
Figure \ref{fig:comp_highSFR} shows that many local star-forming
galaxies with high SFRs are able to have
[O\emissiontype{III}]/[O\emissiontype{II}] comparable to the
[O\emissiontype{II}] emitters at $z\sim1.5$, suggesting that high
star-formation activity is important for the high ionization state.  
However, local star-forming galaxies with high SFRs have a wide range
of  [O\emissiontype{III}]/[O\emissiontype{II}] ratios, so high star
formation activity does not necessarily result in a high ionization
state. The middle panel of Figure \ref{fig:comp_highSFR} implies that
star-forming galaxies with high
[O\emissiontype{III}]/[O\emissiontype{II}] ratios $>1$ have different
metallicity at $z\sim0$ and 1.5. Moreover, at a given stellar mass,
there are only a few local galaxies that have high
[O\emissiontype{III}]/[O\emissiontype{II}] ratios comparable to the
[O\emissiontype{II}] emitters at $z\sim1.5$. 

As we probe higher redshifts, closer to the formation epoch of
the galaxies, the ages of galaxies should be
younger. O-type stars (age$<$1--10Myr) as well as Wolf-Rayet stars
(age$\sim$3--5Myr) may  be present. Also, \citet{Stanway2014}
shows that high [O\emissiontype{III}]/H$\beta$ ratios with high
ionization parameter can be reproduced for up to $\sim$100 Myr by
taking into account binary stars, which have a significant impact on
the evolution of massive and Wolf-Rayet stars. Therefore, 
one possibility is that massive young stellar populations harden the
radiation field of high-$z$ galaxies.

\subsection{Comparison with other high-$z$ galaxies}
\label{sec:comparison-other-studies}

\citet{Stott2013} studied 193 H$\alpha$ emitters at $z=0.84$ and 1.47,
which were selected from the High-redshift Emission Line Survey
(HiZELS). The observation was conducted in high resolution mode with
Subaru/FMOS. Among the H$\alpha$ emitters, 152 galaxies are at $z=1.47$.
\citet{Zahid2014} studied 162 star-forming galaxies at $z\sim1.6$
in the COSMOS field, which were also observed in high resolution mode
with Subaru/FMOS. Among the galaxies with H$\alpha$ detected in $H$-band
spectra, 87 galaxies were observed in $J$-band. 
\citet{Yabe2014} studied 343 star-forming galaxies at $z\sim1.4$
in the SXDS/UDS field, which were observed in low-resolution mode with
Subaru/FMOS. These studies have detected H$\alpha$, H$\beta$,
[O\emissiontype{III}]$\lambda5007$, and
[N\emissiontype{II}]$\lambda6584$ lines for individual galaxies and 
stacked spectra, and then investigated the line ratios and gas-phase
metallicity. The number of galaxies in these studies is a factor of
2--3 larger than our sample. However, our study is the only one
with individual intensity measurements of [O\emissiontype{II}] for 118
galaxies at $z\approx1.47$ and 1.62.

In terms of BPT diagnostics, we note that our result is
consistent with previous high-$z$ studies. Thus, there is a consensus
that star-forming galaxies at $z\approx1.4$--1.6 are offset from local
star-forming galaxies with different [O\emissiontype{III}]/H$\beta$
and [N\emissiontype{II}]/H$\alpha$ line ratios (Figure \ref{fig:BPT}).

There is a discrepancy between
the stellar mass--[N\emissiontype{II}]/H$\alpha$ relation that we 
measure and those of \citet{Zahid2014} and
\citet{Yabe2014}. At a given stellar mass, star-forming galaxies in our
sample show larger [N\emissiontype{II}]/H$\alpha$ ratios (Figure \ref{fig:N2}). 
Note that we convert all stellar mass estimates to a common IMF \citep{Chabrier2003},
where applicable. Therefore, the reason for the disagreement is unclear.
One possibility is a selection bias. 
Our galaxies are selected by their [O\emissiontype{II}] emission,
while these studies selected by their rest-frame optical continuum
($K_s$-selected).  
\citet{Stott2013} find a higher [N\emissiontype{II}]/H$\alpha$ ratio
as a function of stellar mass. They had selected galaxies by their
H$\alpha$ emission.
While our result is consistent with that of \citet{Stott2013} at
stellar mass $\gtrsim10^{10}$ \MO, the
[N\emissiontype{II}]/H$\alpha$ ratios of our galaxies are lower at
stellar mass $\lesssim10^{10}$ \MO.   
Since the sample used in \citet{Stott2013} includes galaxies at 
$z=0.84$, it may result in higher line ratios, due to the
evolving mass--metallicity relation.
It is also worth considering whether the ionization parameter and/or
N/O abundance are affected by selection effects. However, this
investigation requires additional observations, as previous studies do
not have information on the [O\emissiontype{II}] emission.

Large NIR spectroscopic surveys are being conducted at higher
redshifts ($z\approx2.0$--2.6) with Keck/MOSFIRE. 
\citet{Steidel2014} presented initial results of the Keck Baryonic
Structure Survey which studies 179 star-forming galaxies at
$z\sim2.3$. \citet{Shapley2014} presented early observations of the
MOSFIRE Deep Evolution Field survey, which studies 118
star-forming galaxies at $z\sim2.3$. They also show that there is an
offset on the BPT diagram from local galaxies \citep[Figure
\ref{fig:BPT} and see also e.g.,][]{Erb2006a}, and that the N/O
abundance ratio of galaxies at $z\sim2.3$ is likely to be different
from local galaxies. Therefore, strong [O\emissiontype{III}] emission
and high N/O ratio appear to be common in high-$z$ galaxies at
$z\approx$ 1.5--2.5.

\section{Conclusions}
\label{sec:conclusions}

We conducted near-infrared spectroscopy with FMOS on the Subaru
Telescope for 118 [O\emissiontype{II}] emission-line galaxies at
$z\sim1.5$ in the Subaru Deep Field, to investigate the physical state of
the interstellar medium of typical star-forming galaxies at high
redshift, using rest-frame optical nebular emission lines.
The galaxy sample consists of [O\emissiontype{II}] emitters at
$z\approx$1.47 and 1.62 selected by NB921 and NB973 narrow-band
imaging with Suprime-Cam on the Subaru Telescope
\citep{Ly2007,Ly2012}. Moderate-resolution ($R\sim2200$) spectra in two
wavelength regions, 1.11--1.35$\mu$m and 1.60--1.80$\mu$m,
were obtained.

We have detected the H$\alpha$ emission line in $H$-band spectra for 115
galaxies, and [O\emissiontype{III}]$\lambda$5007 emission line in $J$-band
spectra for 45 galaxies. Among these galaxies, H$\beta$,
[N\emissiontype{II}]$\lambda$6584,
[S\emissiontype{II}]$\lambda\lambda$6716,6731 are also detected for 13, 16,
and 6 galaxies, respectively. Including the
[O\emissiontype{II}]$\lambda$3727 emission line measured by
narrow-band imaging, we use six major nebular emission lines in
the rest-frame optical to investigate physical state of the interstellar 
medium in typical star-forming galaxies at $z\sim$1.5.

We have found a tight correlation between H$\alpha$ and
[O\emissiontype{II}], which suggests that [O\emissiontype{II}] can be
a good indicator of star formation activity for galaxies at
$z\sim1.5$. The line ratios of H$\alpha$/[O\emissiontype{II}] are
consistent with those of local galaxies. We have also found that 
the  H$\alpha$/[O\emissiontype{II}] line ratio is dependent on
ionization parameter and stellar mass. The [O\emissiontype{II}]
emitters suffer dust extinction of $\langle$A(H$\alpha$)$\rangle$=0.63
mag, which is estimated from the H$\alpha$/H$\beta$ Balmer decrement.
 
We have found that [O\emissiontype{II}] emitters have strong
[O\emissiontype{III}] emission. 
The [O\emissiontype{III}]/[O\emissiontype{II}] ratios are larger than
normal star-forming galaxies in the local Universe, suggesting that 
ionization parameter is enhanced for typical star-forming galaxies
at $z\sim1.5$. Less massive galaxies have larger
[O\emissiontype{III}]/[O\emissiontype{II}] ratios.
With the result that the electron densities in our galaxies are
consistent with local galaxies, the high ionization parameter of
galaxies at high redshifts may be attributed to a harder radiation
field by young stellar populations and/or the increasing number of
ionizing photon from massive stars. 

At a given oxygen abundance, star-forming galaxies at $z\sim1.5$ may
have relatively higher nitrogen-to-oxygen abundance ratio, compared to
local galaxies. However, no clear difference in N/O abundance ratio is
seen between the galaxies at $z\sim1.5$ and $z\sim0$ at a given
stellar mass. These contrasting results suggest that the evolution of
mass-metallicity relation is responsible for the observed trends in
N/O abundance ratio at $z\sim1.5$. 
The differences in the physical state of the ISM from
local galaxies indicate that direct measurements of metallicity, not
using diagnostics with strong emission lines which are calibrated with
local galaxies, is required to properly study the mass-metallicity
relation at high redshifts.

\section{Acknowledgments}
\noindent
We thank Enrique P$\acute{\rm e}$rez-Montero for reviewing our
manuscript and providing helpful comments, which improved the paper.  
The data used in this paper were collected at the Subaru Telescope,
which is operated by the National Astronomical Observatory of Japan. 
We thank the Subaru Telescope staff for their invaluable help in
assisting our observations. MH acknowledges support from the Japan
Society for the Promotion of Science (JSPS) through JSPS Research
Fellowship for Young Scientists.
CL is funded through the NASA Postdoctoral Program.

\bigskip


\clearpage
\begin{landscape}
  \begin{deluxetable}{lcccccccccc}
  \tabletypesize{\normalsize}
  \tablewidth{0pc}
  \tablecaption{[O\emissiontype{II}] emitters in the SDF that are confirmed by Subaru/FMOS spectroscopy.}
  \tablehead{
  \colhead{ID} &
  \colhead{R.A.} &
  \colhead{Dec.} &
  \colhead{$z_{\rm spec}$\tablenotemark{\P}} &
  \colhead{[O\emissiontype{II}]$\lambda$3727\tablenotemark{\dagger}} &
  \colhead{H$\beta$$^\ddagger$} &
  \colhead{[O\emissiontype{III}]$\lambda$5007\tablenotemark{\ddagger}} &
  \colhead{H$\alpha$$^\ddagger$} &
  \colhead{[N\emissiontype{II}]$\lambda$6584\tablenotemark{\ddagger}} &
  \colhead{[S\emissiontype{II}]$\lambda\lambda$6716,6731\tablenotemark{\ddagger}}}

\startdata
NB921$\_$20415  & 201.008072 & +27.211218 & 1.465 &  2.12$\pm$0.05  & ---           & ---            &  2.63$\pm$0.31  & \udots          & \udots \\
NB921$\_$20750  & 201.080215 & +27.211807 & 1.457 &  0.72$\pm$0.05  & \udots        & \udots         &  1.14$\pm$0.34  & \udots          & \udots \\
NB921$\_$21503  & 200.954636 & +27.213587 & 1.465 &  0.75$\pm$0.05  & ---           & ---            &  1.46$\pm$0.31  & \udots          & \udots \\
NB921$\_$26406  & 201.117203 & +27.227135 & 1.474 &  1.32$\pm$0.07  & ---           & ---            &  2.70$\pm$0.80  & \udots          & \udots \\
NB921$\_$28540  & 201.023178 & +27.231798 & 1.460 &  0.75$\pm$0.05  & ---           & ---            &  1.63$\pm$0.42  & \udots          & \udots \\
NB921$\_$41594  & 201.143753 & +27.269798 & 1.464 &  1.00$\pm$0.04  & \udots        & \udots         &  1.61$\pm$0.28  & \udots          & \udots \\
NB921$\_$42750  & 201.172729 & +27.270859 & 1.488 &  1.98$\pm$0.14  & ---           & ---            &  1.88$\pm$0.23  & \udots          & \udots \\
NB921$\_$42929  & 201.018433 & +27.271896 & 1.464 &  1.83$\pm$0.05  & \udots        &  6.16$\pm$1.61 &  3.85$\pm$0.21  & \udots          & \udots \\
NB921$\_$49590  & 201.027985 & +27.292707 & 1.472 &  2.58$\pm$0.06  & ---           & ---            &  3.87$\pm$0.45  &  1.24$\pm$0.35  & \udots \\
NB921$\_$51600  & 201.129227 & +27.295822 & 1.469 &  2.02$\pm$0.05  & \udots        &  1.84$\pm$0.34 &  3.04$\pm$0.28  & \udots          & \udots \\
NB921$\_$52455  & 201.111649 & +27.297915 & 1.473 &  1.62$\pm$0.05  & \udots        &  6.18$\pm$0.28 &  2.32$\pm$0.29  & \udots          & \udots \\
NB921$\_$53008  & 201.323746 & +27.299255 & 1.484 &  1.27$\pm$0.08  & ---           & ---            &  1.70$\pm$0.10  & \udots          & \udots \\
NB921$\_$53257  & 200.942825 & +27.301119 & 1.477 &  3.98$\pm$0.03  & 3.12$\pm$0.01 & 13.41$\pm$0.19 &  8.91$\pm$0.23  &  0.82$\pm$0.21  &  0.75$\pm$0.14,  0.63$\pm$0.13 \\
NB921$\_$53590  & 201.292923 & +27.300709 & 1.462 &  0.62$\pm$0.04  & \udots        & \udots         &  1.63$\pm$0.18  & \udots          & \udots \\
NB921$\_$53725  & 200.904846 & +27.303421 & 1.476 &  4.76$\pm$0.11  & \udots        & \udots         &  3.40$\pm$0.54  & \udots          & \udots \\
NB921$\_$54675  & 200.977539 & +27.304607 & 1.462 &  1.25$\pm$0.13  & \udots        & \udots         &  3.27$\pm$0.42  &  2.15$\pm$0.42  & \udots \\
NB921$\_$57905  & 201.268173 & +27.317118 & 1.481 &  2.92$\pm$0.10  & \udots        &  6.60$\pm$0.90 &  2.97$\pm$0.30  & \udots          & \udots \\
NB921$\_$59540  & 200.910477 & +27.321886 & 1.465 &  1.47$\pm$0.05  & \udots        & \udots         &  1.59$\pm$0.22  & \udots          & \udots \\
NB921$\_$61282  & 201.276123 & +27.327734 & 1.483 &  1.12$\pm$0.04  & 1.53$\pm$0.35 &  1.73$\pm$0.37 &  2.84$\pm$0.13  & \udots          & \udots \\
NB921$\_$62719  & 201.008759 & +27.332539 & 1.466 &  1.42$\pm$0.05  & \udots        & \udots         &  3.97$\pm$0.56  & \udots          & \udots \\
NB921$\_$63744  & 200.943481 & +27.337433 & 1.481 &  5.36$\pm$0.16  & 2.17$\pm$0.31 &  6.13$\pm$0.81 &  6.22$\pm$0.36  & \udots          & \udots \\
NB921$\_$67247  & 201.047211 & +27.349859 & 1.469 &  0.53$\pm$0.04  & ---           & ---            &  0.94$\pm$0.20  & \udots          & \udots \\
NB921$\_$68313  & 200.997437 & +27.355768 & 1.469 &  1.95$\pm$0.06  & \udots        & \udots         &  1.98$\pm$0.62  & \udots          & \udots \\
NB921$\_$69065  & 201.120758 & +27.358625 & 1.473 &  9.15$\pm$0.09  & \udots        & 10.40$\pm$0.12 & 13.20$\pm$1.22  &  3.88$\pm$0.43  &  1.71$\pm$0.61,  1.13$\pm$0.36 \\
NB921$\_$72036  & 201.153427 & +27.368980 & 1.455 &  1.23$\pm$0.05  & \udots        & \udots         &  2.30$\pm$0.17  & \udots          & \udots \\
NB921$\_$73875  & 200.918503 & +27.376816 & 1.465 &  3.95$\pm$0.13  & \udots        & \udots         &  3.75$\pm$0.39  & \udots          & \udots \\
NB921$\_$74041  & 201.194672 & +27.377974 & 1.456 &  1.69$\pm$0.05  & \udots        &  9.84$\pm$1.60 &  4.53$\pm$0.27  & \udots          & \udots \\
NB921$\_$74192  & 201.194168 & +27.377691 & 1.456 &  0.56$\pm$0.02  & 1.83$\pm$0.31 &  1.89$\pm$0.31 &  1.52$\pm$0.21  & \udots          & \udots \\
NB921$\_$76059  & 201.091171 & +27.385765 & 1.471 &  3.37$\pm$0.10  & \udots        & \udots         & 10.16$\pm$0.92  &  5.11$\pm$0.74  & \udots \\
NB921$\_$80256  & 200.976547 & +27.400311 & 1.465 &  0.48$\pm$0.03  & ---           & ---            &  1.08$\pm$0.32  & \udots          & \udots \\
NB921$\_$81619  & 200.982117 & +27.405230 & 1.466 &  0.46$\pm$0.03  & ---           & ---            &  0.58$\pm$0.15  & \udots          & \udots \\
NB921$\_$84384  & 201.035858 & +27.416105 & 1.464 &  2.80$\pm$0.05  & 1.21$\pm$0.16 &  3.92$\pm$0.20 &  3.47$\pm$0.25  & \udots          & \udots \\
NB921$\_$84670  & 201.296234 & +27.416384 & 1.476 &  6.12$\pm$0.25  & \udots        & \udots         &  6.31$\pm$1.07  & \udots          & \udots \\
NB921$\_$87249  & 201.267166 & +27.426624 & 1.456 &  2.98$\pm$0.08  & \udots        & \udots         &  7.00$\pm$0.53  &  2.02$\pm$0.28  & \udots \\
NB921$\_$87748  & 201.132797 & +27.429152 & 1.471 &  3.21$\pm$0.06  & \udots        &  5.80$\pm$1.50 &  6.47$\pm$0.51  &  1.02$\pm$0.28  &  1.11$\pm$0.27,  0.99$\pm$0.28 \\
NB921$\_$90789  & 201.159622 & +27.440001 & 1.487 &  1.05$\pm$0.08  & 1.37$\pm$0.07 &  4.30$\pm$0.20 &  3.92$\pm$0.16  & \udots          & \udots \\
NB921$\_$90839  & 200.900070 & +27.439596 & 1.453 &  0.73$\pm$0.05  & \udots        & \udots         &  1.63$\pm$0.45  & \udots          & \udots \\
NB921$\_$91027  & 201.097717 & +27.441690 & 1.472 &  2.33$\pm$0.14  & ---           & ---            &  1.54$\pm$0.46  & \udots          & \udots \\
NB921$\_$91549  & 201.068466 & +27.442301 & 1.470 &  2.36$\pm$0.09  & \udots        & \udots         &  4.21$\pm$1.39  & \udots          & \udots \\
NB921$\_$95558  & 200.978973 & +27.457714 & 1.476 &  2.66$\pm$0.10  & \udots        & \udots         &  2.24$\pm$0.66  & \udots          & \udots \\
NB921$\_$96152  & 200.954590 & +27.462772 & 1.477 &  2.07$\pm$0.08  & \udots        & \udots         &  2.40$\pm$0.33  & \udots          & \udots \\
NB921$\_$97477  & 201.216415 & +27.464403 & 1.485 &  0.56$\pm$0.06  & 0.77$\pm$0.10 &  3.38$\pm$0.14 &  1.61$\pm$0.16  &  0.36$\pm$0.07  & \udots \\
NB921$\_$101385 & 200.912842 & +27.481104 & 1.466 &  2.19$\pm$0.06  & \udots        & \udots         &  3.34$\pm$0.41  & \udots          & \udots \\
NB921$\_$102638 & 201.060425 & +27.486040 & 1.472 &  1.03$\pm$0.05  & ---           & ---            &  2.20$\pm$0.41  & \udots          & \udots \\
NB921$\_$103150 & 201.024399 & +27.487307 & 1.481 &  1.20$\pm$0.07  & \udots        & \udots         &  1.74$\pm$0.19  & \udots          & \udots \\
NB921$\_$103705 & 201.003967 & +27.490063 & 1.475 & 13.88$\pm$0.48  & \udots        & 22.12$\pm$4.73 & 16.96$\pm$2.02  &  9.29$\pm$1.81  &  3.21$\pm$1.09,  4.60$\pm$1.13 \\
NB921$\_$104655 & 201.087936 & +27.493420 & 1.464 &  1.22$\pm$0.03  & \udots        &  6.37$\pm$0.15 &  3.40$\pm$0.29  &  0.61$\pm$0.15  & \udots \\
NB921$\_$105983 & 201.114471 & +27.500454 & 1.485 &  1.29$\pm$0.10  & \udots        &  3.93$\pm$0.31 &  2.87$\pm$0.39  & \udots          & \udots \\
NB921$\_$107118 & 201.019028 & +27.504000 & 1.475 &  1.54$\pm$0.07  & \udots        & \udots         &  3.38$\pm$0.57  & \udots          & \udots \\
NB921$\_$109072 & 200.994446 & +27.512560 & 1.473 &  5.21$\pm$0.18  & \udots        & \udots         &  5.93$\pm$0.98  & \udots          & \udots \\
NB921$\_$110460 & 201.324646 & +27.516947 & 1.476 &  3.92$\pm$0.12  & \udots        &  4.20$\pm$0.85 &  4.82$\pm$0.86  & \udots          & \udots \\
NB921$\_$110480 & 201.070343 & +27.517275 & 1.463 &  1.61$\pm$0.02  & 1.59$\pm$0.29 &  5.24$\pm$0.93 &  4.44$\pm$0.16  &  0.27$\pm$0.09  & \udots \\
NB921$\_$110936 & 201.257339 & +27.518309 & 1.477 &  0.51$\pm$0.03  & 1.12$\pm$0.22 &  1.88$\pm$0.27 &  1.73$\pm$0.15  & \udots          & \udots \\
NB921$\_$113063 & 201.060104 & +27.529142 & 1.475 &  6.15$\pm$0.18  & ---           & ---            &  2.32$\pm$0.51  & \udots          & \udots \\
NB921$\_$114814 & 201.170425 & +27.533533 & 1.475 &  1.55$\pm$0.05  & \udots        & \udots         &  2.37$\pm$0.56  & \udots          & \udots \\
NB921$\_$116328 & 201.170258 & +27.538916 & 1.476 &  0.99$\pm$0.04  & \udots        &  3.50$\pm$0.18 & \udots          & \udots          & \udots \\
NB921$\_$116643 & 201.074005 & +27.541149 & 1.481 &  1.54$\pm$0.11  & \udots        & \udots         &  3.01$\pm$0.13  & \udots          & \udots \\
NB921$\_$119072 & 201.065796 & +27.551952 & 1.463 &  3.40$\pm$0.06  & \udots        & \udots         &  8.64$\pm$0.68  &  2.03$\pm$0.50  & \udots \\
NB921$\_$119235 & 201.058258 & +27.550882 & 1.473 &  0.54$\pm$0.04  & \udots        &  2.64$\pm$0.26 & \udots          & \udots          & \udots \\
NB921$\_$119758 & 201.121780 & +27.553385 & 1.468 &  1.47$\pm$0.02  & 2.50$\pm$0.21 & 12.42$\pm$0.54 &  4.44$\pm$0.31  & \udots          & \udots \\
NB921$\_$121008 & 201.245834 & +27.557402 & 1.482 &  1.38$\pm$0.11  & \udots        &  9.59$\pm$0.86 &  2.63$\pm$0.12  & \udots          & \udots \\
NB921$\_$122309 & 201.245865 & +27.564041 & 1.461 &  2.01$\pm$0.08  & \udots        & \udots         &  3.37$\pm$0.49  & \udots          & \udots \\
NB921$\_$126941 & 201.247879 & +27.579729 & 1.462 &  1.96$\pm$0.06  & \udots        & \udots         &  4.46$\pm$0.55  & \udots          & \udots \\
NB921$\_$127005 & 200.998505 & +27.580166 & 1.454 &  2.29$\pm$0.08  & \udots        & \udots         &  3.40$\pm$0.31  &  0.58$\pm$0.17  & \udots \\
NB921$\_$127149 & 201.096802 & +27.580116 & 1.478 &  0.86$\pm$0.07  & ---           & ---            &  1.23$\pm$0.40  & \udots          & \udots \\
NB921$\_$129002 & 200.981079 & +27.589703 & 1.454 &  0.77$\pm$0.04  & \udots        & \udots         &  2.61$\pm$0.45  & \udots          & \udots \\
NB921$\_$132078 & 201.049194 & +27.599440 & 1.475 &  1.94$\pm$0.07  & \udots        & \udots         &  2.76$\pm$0.57  & \udots          & \udots \\
NB921$\_$133042 & 201.080048 & +27.603086 & 1.479 &  0.80$\pm$0.07  & \udots        & \udots         &  1.34$\pm$0.36  & \udots          & \udots \\
NB921$\_$133516 & 201.124435 & +27.609943 & 1.473 &  2.20$\pm$0.02  & 2.60$\pm$0.35 &  2.84$\pm$0.71 &  7.38$\pm$0.50  &  2.80$\pm$0.23  & \udots \\
NB921$\_$133703 & 201.124420 & +27.611273 & 1.476 & 11.95$\pm$0.15  & \udots        & 10.61$\pm$1.02 & 14.98$\pm$0.60  &  1.91$\pm$0.32  &  1.71$\pm$0.30,  1.31$\pm$0.32 \\
NB921$\_$137733 & 201.037720 & +27.619905 & 1.482 &  0.43$\pm$0.06  & ---           & ---            &  0.90$\pm$0.25  & \udots          & \udots \\
NB921$\_$139980 & 201.193390 & +27.629015 & 1.475 &  0.98$\pm$0.08  & ---           & ---            &  1.56$\pm$0.48  & \udots          & \udots \\
NB921$\_$140463 & 201.129898 & +27.631865 & 1.478 &  3.48$\pm$0.08  & \udots        &  7.68$\pm$0.18 &  4.64$\pm$0.28  & \udots          &  0.59$\pm$0.21,  0.65$\pm$0.21 \\
NB921$\_$148013 & 201.166000 & +27.658577 & 1.476 &  5.18$\pm$0.09  & \udots        &  5.41$\pm$0.38 &  4.75$\pm$0.50  & \udots          & \udots \\
NB921$\_$154909 & 201.143539 & +27.681538 & 1.475 &  0.49$\pm$0.04  & ---           & ---            &  0.79$\pm$0.21  & \udots          & \udots \\
NB921$\_$155168 & 201.056686 & +27.683475 & 1.477 &  2.00$\pm$0.06  & \udots        & \udots         &  2.42$\pm$0.52  & \udots          & \udots \\
NB921$\_$155600 & 201.251480 & +27.684908 & 1.479 &  4.60$\pm$0.11  & \udots        & \udots         &  6.97$\pm$0.76  & \udots          & \udots \\
NB921$\_$159865 & 201.242142 & +27.698488 & 1.477 &  1.57$\pm$0.08  & \udots        & \udots         &  1.51$\pm$0.47  & \udots          & \udots \\
NB921$\_$175932 & 201.045105 & +27.753534 & 1.466 &  0.48$\pm$0.03  & ---           & ---            &  0.69$\pm$0.17  & \udots          & \udots \\
NB973$\_$39461  & 201.021744 & +27.253485 & 1.589 &  2.92$\pm$0.28  & \udots        & \udots         &  3.35$\pm$0.18  & \udots          & \udots \\
NB973$\_$53162  & 201.155777 & +27.293501 & 1.589 &  7.87$\pm$0.59  & \udots        & 12.54$\pm$0.83 &  4.96$\pm$0.56  & \udots          & \udots \\
NB973$\_$53375  & 201.108231 & +27.291931 & 1.595 &  0.47$\pm$0.13  & \udots        & \udots         &  1.03$\pm$0.34  & \udots          & \udots \\
NB973$\_$53557  & 200.968735 & +27.293095 & 1.621 &  0.78$\pm$0.10  & \udots        &  3.34$\pm$0.34 &  1.26$\pm$0.29  & \udots          & \udots \\
NB973$\_$57024  & 201.082626 & +27.301146 & 1.589 &  0.86$\pm$0.23  & \udots        &  1.32$\pm$0.23 &  1.49$\pm$0.31  & \udots          & \udots \\
NB973$\_$63423  & 201.185745 & +27.323143 & 1.590 &  2.35$\pm$0.33  & \udots        & \udots         &  2.58$\pm$0.53  & \udots          & \udots \\
NB973$\_$63500  & 201.184814 & +27.322449 & 1.588 &  2.21$\pm$0.27  & ---           & ---            &  4.97$\pm$0.41  & \udots          & \udots \\
NB973$\_$65046  & 201.151443 & +27.328642 & 1.599 &  2.07$\pm$0.25  & \udots        & \udots         &  4.33$\pm$0.57  & \udots          & \udots \\
NB973$\_$65525  & 201.297684 & +27.330811 & 1.616 &  0.37$\pm$0.09  & ---           & ---            &  0.96$\pm$0.13  & \udots          & \udots \\
NB973$\_$69720  & 201.184006 & +27.351362 & 1.616 &  3.05$\pm$0.24  & \udots        &  5.59$\pm$0.49 &  5.02$\pm$0.43  & \udots          & \udots \\
NB973$\_$71489  & 200.983398 & +27.361053 & 1.590 &  4.01$\pm$0.38  & \udots        & \udots         &  4.31$\pm$0.77  & \udots          & \udots \\
NB973$\_$72086  & 200.994644 & +27.363752 & 1.590 &  2.11$\pm$0.29  & \udots        &  3.04$\pm$0.21 &  2.19$\pm$0.33  & \udots          & \udots \\
NB973$\_$72650  & 201.210007 & +27.365982 & 1.600 &  1.83$\pm$0.14  & \udots        & \udots         &  6.13$\pm$0.56  & \udots          & \udots \\
NB973$\_$72934  & 201.216019 & +27.368015 & 1.599 &  5.08$\pm$0.37  & \udots        & \udots         &  5.06$\pm$0.49  & \udots          & \udots \\
NB973$\_$79083  & 201.158401 & +27.401751 & 1.588 &  1.56$\pm$0.27  & \udots        &  4.29$\pm$0.22 &  2.53$\pm$0.27  & \udots          & \udots \\
NB973$\_$81424  & 201.315735 & +27.412804 & 1.591 &  3.79$\pm$0.31  & \udots        &  6.83$\pm$1.01 &  5.71$\pm$0.90  & \udots          & \udots \\
NB973$\_$85096  & 200.967087 & +27.428595 & 1.599 &  0.74$\pm$0.15  & ---           & ---            &  1.23$\pm$0.23  & \udots          & \udots \\
NB973$\_$90084  & 201.320602 & +27.450769 & 1.591 &  0.81$\pm$0.21  & ---           & ---            &  1.24$\pm$0.29  & \udots          & \udots \\
NB973$\_$90805  & 201.290176 & +27.456099 & 1.613 &  1.93$\pm$0.23  & \udots        & \udots         &  2.54$\pm$0.34  &  0.96$\pm$0.22  & \udots \\
NB973$\_$91278  & 200.929718 & +27.457098 & 1.597 &  1.84$\pm$0.20  & \udots        & \udots         &  2.68$\pm$0.34  & \udots          & \udots \\
NB973$\_$91531  & 201.048340 & +27.458298 & 1.632 &  0.73$\pm$0.14  & \udots        &  3.65$\pm$0.14 & \udots          & \udots          & \udots \\
NB973$\_$104691 & 201.186508 & +27.529871 & 1.584 &  0.92$\pm$0.24  & \udots        &  3.51$\pm$0.24 &  2.39$\pm$0.33  & \udots          & \udots \\
NB973$\_$104719 & 200.941956 & +27.531408 & 1.617 &  1.88$\pm$0.06  & 1.63$\pm$0.56 &  3.50$\pm$0.80 &  4.02$\pm$0.21  & \udots          & \udots \\
NB973$\_$105584 & 201.206970 & +27.536465 & 1.606 &  2.66$\pm$0.20  & \udots        &  6.47$\pm$0.82 &  5.22$\pm$0.43  & \udots          & \udots \\
NB973$\_$108859 & 201.048325 & +27.549301 & 1.618 &  1.84$\pm$0.15  & \udots        &  4.71$\pm$0.27 &  4.15$\pm$0.57  & \udots          & \udots \\
NB973$\_$109063 & 200.925018 & +27.550058 & 1.616 &  2.57$\pm$0.22  & \udots        & \udots         &  5.17$\pm$1.25  & \udots          & \udots \\
NB973$\_$109503 & 200.923508 & +27.552755 & 1.599 &  2.54$\pm$0.19  & \udots        &  7.54$\pm$0.25 &  6.30$\pm$0.69  & \udots          & \udots \\
NB973$\_$115023 & 201.158646 & +27.577322 & 1.644 &  0.45$\pm$0.12  & 0.76$\pm$0.15 &  1.60$\pm$0.27 &  1.86$\pm$0.38  & \udots          & \udots \\
NB973$\_$116135 & 201.195541 & +27.582874 & 1.589 &  1.86$\pm$0.31  & \udots        & \udots         &  3.38$\pm$0.73  & \udots          & \udots \\
NB973$\_$117371 & 201.193390 & +27.589464 & 1.585 &  1.69$\pm$0.38  & ---           & ---            &  1.66$\pm$0.51  & \udots          & \udots \\
NB973$\_$118086 & 201.349854 & +27.592388 & 1.596 &  0.67$\pm$0.14  & \udots        &  2.35$\pm$0.38 &  1.08$\pm$0.30  & \udots          & \udots \\
NB973$\_$123696 & 200.982178 & +27.620754 & 1.592 &  3.12$\pm$0.28  & \udots        & \udots         &  4.61$\pm$0.85  & \udots          & \udots \\
NB973$\_$124743 & 200.944931 & +27.624931 & 1.580 &  4.44$\pm$0.80  & \udots        & \udots         &  1.90$\pm$0.45  & \udots          & \udots \\
NB973$\_$128025 & 201.226929 & +27.641090 & 1.610 &  1.77$\pm$0.22  & \udots        & \udots         &  2.16$\pm$0.64  & \udots          & \udots \\
NB973$\_$128071 & 201.285156 & +27.641356 & 1.589 &  1.43$\pm$0.26  & ---           & ---            &  2.15$\pm$0.45  & \udots          & \udots \\
NB973$\_$131961 & 201.095764 & +27.658642 & 1.585 &  2.05$\pm$0.38  & \udots        & \udots         &  2.74$\pm$0.46  & \udots          & \udots \\
NB973$\_$133848 & 201.094513 & +27.665018 & 1.603 &  0.58$\pm$0.14  & \udots        & \udots         &  1.40$\pm$0.43  & \udots          & \udots \\
NB973$\_$136464 & 201.200027 & +27.675671 & 1.604 &  2.19$\pm$0.20  & \udots        &  3.66$\pm$0.38 &  4.13$\pm$1.18  & \udots          & \udots \\
NB973$\_$147549 & 201.049255 & +27.726721 & 1.598 &  1.38$\pm$0.17  & \udots        &  5.14$\pm$0.78 &  2.74$\pm$0.34  & \udots          & \udots \\
\enddata
\vspace{-0.75cm}
\label{tab:individual_obj}
\tablecomments{Galaxies without $J$-band spectrum available are marked with ---, and non-detections are indicated with \udots.}
\tablenotetext{\P}{Average of the redshifts measured from detected emission lines.}
\tablenotetext{\dagger}{The fluxes are measured from narrow-band imaging, and corrected for the filter response function as well as dust extinction. Fluxes are in units of $10^{-16}$ erg s$^{-1}$ cm$^{-2}$.}
\tablenotetext{\ddagger}{The fluxes of emission line are all corrected for dust extinction (see the text in \S \ref{sec:Balmer_decrement} for the detail). Fluxes are in units of $10^{-16}$ erg s$^{-1}$ cm$^{-2}$.}

\end{deluxetable}
\clearpage
\end{landscape}


\begin{deluxetable}{cccrccccccc}
  \tabletypesize{\normalsize}
  \tablewidth{0pc}
  \tablecaption{Subsamples for spectral stacking, which are divided based on stellar mass.}
  \tablehead{
    \colhead{} &
    \colhead{Coverage} &
    \colhead{N$_{\rm total}$} &
    \colhead{Subsample} &
    \colhead{N} &
    \colhead{$\log(M_\star/\Mo)$} &
    \colhead{H$\beta$} &
    \colhead{[O\emissiontype{III}]} &
    \colhead{H$\alpha$} &
    \colhead{[N\emissiontype{II}]} &
    \colhead{[S\emissiontype{II}]}}
  \startdata
  sample-1 & $J$ and $H$ & 89  & & 89 & 9.26 [8.49, 11.2] & 13 & 43 & 86 & 14 & 5 \\
  \hline
  \multirow{3}{*}{sample-2}  & \multirow{3}{*}{$J$ and $H$} &
  \multirow{3}{*}{89} & 2a ...... &9  & 8.83 [8.49, 9.00] & 2 &  7 &  7 & 1 & 0   \\
  & &         &2b ...... & 68 & 9.62 [9.00, 10.0] & 9 & 30 & 67 & 8 & 3   \\
  & &         &2c ...... & 12 & 10.3 [10.0, 11.2] & 2 &  6 & 12 & 5 & 2   \\
  \hline
  \multirow{4}{*}{sample-3} & \multirow{4}{*}{$H$}  &
  \multirow{4}{*}{113} &3a ...... &  47 & 9.25 [8.49, 9.45] & --- & --- & 44 & 3 & 1   \\
  & &   &3b ...... &  30 & 9.46 [9.45, 9.70] & --- & --- & 30 & 3 & 1   \\
  & &   &3c ...... &  22 & 9.85 [9.70, 10.0] & --- & --- & 22 & 4 & 1   \\
  & &   &3d ...... &  14 & 10.3 [10.0, 11.2] & --- & --- & 14 & 5 & 2   \\
  \enddata
  \vspace{-0.5cm}
  \label{tab:stack_subsample}
  \tablecomments{The median stellar masses are shown in each subsamples, and the
    values in brackets are the minimum and maximum of stellar mass in each
    subsamples. Sample-1 and 2 contain the [O\emissiontype{II}] emitters that are
    spectroscopically confirmed and have FMOS spectra in both $J$- and $H$-band.
    Sample-3 is the same as the others, but the galaxies with FMOS spectra in
    $H$-band available are all included. The number shown is that of individual
    emission lines detected in each subsample. Each stacked spectrum is shown
    in Figure \ref{fig:stackedspectra}.}
\end{deluxetable}


\begin{deluxetable}{crccccccc}
  \tabletypesize{\normalsize}
  \tablewidth{0pc}
  \tablecaption{Intensity ratios of emission lines for the stacked spectra of
    individual subsamples which are summarized in Table
    \ref{tab:stack_subsample}. All line ratios are corrected for
    dust extinction, except for H$\alpha$/H$\beta$.}  
  \tablehead{
    \colhead{} & 
    \colhead{Subsample} &
    \colhead{[O\emissiontype{III}]/[O\emissiontype{II}]\tablenotemark{a}} &  
    \colhead{[O\emissiontype{III}]/H$\beta$\tablenotemark{b}} &  
    \colhead{$R_{23}$\tablenotemark{c}} &  
    \colhead{H$\alpha$/H$\beta$\tablenotemark{d}} & 
    \colhead{[N\emissiontype{II}]/H$\alpha$\tablenotemark{e}} &  
    \colhead{[N\emissiontype{II}]/[O\emissiontype{II}]\tablenotemark{f}} &  
    \colhead{[S\emissiontype{II}] doublet\tablenotemark{g}}
  }
  \startdata
    sample-1 & & 1.80$\pm$0.02 & 2.97$\pm$0.06 & 5.61$\pm$0.11 & 3.53$\pm$0.07 & 0.16$\pm$0.01 & 0.28$\pm$0.01 & 1.47$\pm$0.08 \\ 
    \hline
    \multirow{3}{*}{sample-2} & 2a ...... & 4.49$\pm$0.19 & 4.61$\pm$0.26 & 7.17$\pm$0.40 & 3.12$\pm$0.18
    & 0.13$\pm$0.02 & 0.37$\pm$0.06 & 1.74$\pm$0.43 \\
    & 2b ...... & 1.77$\pm$0.02 & 2.98$\pm$0.07 & 5.65$\pm$0.13 & 3.51$\pm$0.08
    & 0.14$\pm$0.01 & 0.23$\pm$0.01 & 1.54$\pm$0.11 \\
    & 2c ...... & 1.60$\pm$0.05 & 1.98$\pm$0.09 & 3.88$\pm$0.17 & 3.53$\pm$0.15
    & 0.34$\pm$0.01 & 0.79$\pm$0.03 & 1.30$\pm$0.11 \\
    \hline
    \multirow{4}{*}{sample-3} & 3a ...... & --- & --- & --- & --- & 0.07$\pm$0.01 & 0.14$\pm$0.02 & 1.19$\pm$0.14 \\
    & 3b ...... & --- & --- & --- & --- & 0.13$\pm$0.01 & 0.23$\pm$0.01 & 2.75$\pm$0.48 \\
    & 3c ...... & --- & --- & --- & --- & 0.24$\pm$0.01 & 0.36$\pm$0.01 & 1.02$\pm$0.11 \\
    & 3d ...... & --- & --- & --- & --- & 0.30$\pm$0.01 & 0.42$\pm$0.01 & 1.45$\pm$0.13 \\
    \enddata
    \label{tab:lineratio}
    \tablenotetext{a}{[O\emissiontype{III}]${\lambda5007}$/[O\emissiontype{II}]${\lambda3727}$}
    \tablenotetext{b}{[O\emissiontype{III}]${\lambda5007}$/H$\beta$}
    \tablenotetext{c}{[O\emissiontype{II}]${\lambda3727}$+[O\emissiontype{III}]${\lambda\lambda4959,5007}$)/H$\beta$}
    \tablenotetext{d}{The observed ratio of H$\alpha$/H$\beta$}
    \tablenotetext{e}{[N\emissiontype{II}]${\lambda6584}$/H$\alpha$}
    \tablenotetext{f}{[N\emissiontype{II}]${\lambda6584}$/[O\emissiontype{II}]${\lambda3727}$}
    \tablenotetext{g}{[S\emissiontype{II}]${\lambda6716}$/[S\emissiontype{II}]${\lambda6731}$}
\end{deluxetable}

\end{document}